\documentclass[a4paper, 11pt]{article}
\usepackage{graphicx}
\usepackage{jheppub}
\usepackage[utf8]{inputenc}
\usepackage{caption}
\usepackage{subcaption}
\usepackage{hyperref}
\usepackage{enumerate}
\usepackage{wrapfig}
\usepackage{amsmath}
\usepackage{amsfonts}
\usepackage{booktabs}
\usepackage{comment}
\usepackage{cancel}
\usepackage{vmargin}
\setmargins{2.5cm}{1.5cm}{16.5cm}{23.42cm}{10pt}{1cm}{0pt}{2cm}
\usepackage{multirow}
\usepackage{slashed}
\usepackage{braket}
\usepackage{amssymb}
\usepackage{setspace}
\usepackage{amssymb}
\usepackage{url}
\usepackage{xcolor}
\usepackage{xspace}
\usepackage{cleveref}
\usepackage{mathabx}

\newcommand{\EOS}{\texttt{EOS}\xspace}
\newcommand{\SUF}{\ensuremath{SU(3)_\mathrm{F}}\xspace}
\newcommand{\nSUF}{\ensuremath{\rm fact. \cancel{\SUF}}\xspace}

\newcommand{\ie}{i.e.\ }
\newcommand{\eg}{e.g.\ }
\newcommand{\pvalue}{$p$-value\xspace}

\title{\boldmath Detailed $SU(3)$ Flavour Symmetry Analysis of Charmless Two-Body $B$-Meson Decays Including Factorizable Corrections}
\author[a,c]{M. Burgos Marcos,}
\emailAdd{m.burgosmarcos@maastrichtuniversity.nl}
\author[b]{M. Reboud,}
\emailAdd{meril.reboud@cnrs.fr}
\author[a,c]{K. K. Vos}
\emailAdd{k.vos@maastrichtuniversity.nl}

\affiliation[a]{Gravitational Waves and Fundamental Physics (GWFP), Maastricht University, Duboisdomein 30, NL-6229 GT Maastricht, the Netherlands}
\affiliation[b]{Université Paris-Saclay, CNRS/IN2P3, IJCLab, 91405 Orsay, France}
\affiliation[c]{Nikhef, Science Park 105, NL-1098 XG Amsterdam, the Netherlands}

\abstract{%
We study the decays of $B_{(s)}$ mesons into light pseudoscalar mesons under the $SU(3)$ flavour symmetry.
Assuming exact $SU(3)$ symmetry at the level of the amplitudes leads to a simple parametrisation.
Using the available experimental data and accounting for mixing effects in the $B_s^0$ branching ratios, we find that the data cannot be described with this assumption.
We improve this parametrisation by including \textit{factorizable} \SUF-breaking effects.
This new approach allows for an excellent description of the data, with a fit $p$ value of $32.3\%$.
We provide posterior predictions for all observables and identify several decay channels that would significantly impact our analysis. 
Finally, we briefly compare our results with the predictions of QCD factorisation, paving the way to a more detailed analysis which could provide insights into QCD effects at low energy scales. 
}

\preprint{%
    EOS-2025-01, Nikhef 2025-004
}

\begin{document}

\maketitle
\newpage
\spacing{1.35}
\section{Introduction}
Non-leptonic two-body charmless $B_{(s)}$ decays have been studied in various approaches.
The plethora of decay possibilities in light mesons and the interference between various decay topologies make these decays especially interesting yet challenging to study.
The experimental programs of both LHCb and the $B$-factories ensure a large amount of experimental information on branching ratios and CP asymmetries.
Especially on the latter, a lot of progress has been made recently.
In 2020, the first observation of time-dependent CP violation in the $B_s^0$ was made by the LHCb collaboration \cite{LHCb:2020byh}, and very recently, Belle II published their first measurement of the direct CP asymmetries of $B^0\to \pi^0\pi^0$ \cite{Belle-II:2024baw}.

Yet, at the same time, the theoretical description of these decays remains challenging.
Several approaches have been explored to predict these observables: theoretical calculations in the framework of QCD factorization \cite{Beneke:1999br,Beneke:2000ry,Beneke:2003zv}, perturbative QCD \cite{Keum:2000ph,Ali:2007ff,Lu:2000em}, lightcone sum rules \cite{Khodjamirian:2000mi, Khodjamirian:2005wn}.
The effect of long-distance final state interactions was discussed in \cite{Atwood:1997iw,Cheng:2004ru}.

Complementary, the decays of $B_{(s)}$ to two pseudoscalar mesons have been studied extensively using flavour symmetries~(see e.g.~\cite{Zeppenfeld:1980ex,Gronau:1990ka,Gronau:1994rj,Gronau:1995hn,Fleischer:1999pa,Buras:1998rb, Buras:2004ub,Gronau:2005kz,Chiang:2004nm,Grossman:2013lya,Fleischer:2007hj} and more recent studies like~\cite{Biswas:2023pyw, Amhis:2022hpm,Bhattacharya:2022akr,Berthiaume:2023kmp,Fleischer:2018bld,Fleischer:2022rkm}).
In addition, several global \SUF-limit analysis, assuming that the QCD interactions of the lightest quarks follow an exact \SUF symmetry, have been 
performed \cite{Cheng:2014rfa,Hsiao:2015iiu,Fu:2003fy,Chiang:2006ih,Chiang:2004nm,Huber:2021cgk,Berthiaume:2023kmp}.\\

In this paper, we present an updated analysis of the $B\to PP$ decays, with $P=\pi, K$, motivated by the updated experimental data. 

First, we perform a global \SUF-limit analysis of the available experimental data.
Compared to some recent global analyses, we include mixing-induced CP asymmetries, we account for experimental correlations between different modes, and we correct for $B_s^0$--$\bar{B}_s^0$ mixing effects in the $B_s^0$ branching ratio \cite{DeBruyn:2012wj}.
We find a poor description of the data.
This discrepancy is partly driven by the $B_s^0\to K^+K^-$ and $B_d^0\to \pi^+\pi^-$ CP asymmetries, where we see these observables alone already indicate a $2\sigma$ tension with the \SUF-limit.

This breakdown of \SUF symmetry can be addressed by introducing \SUF-breaking corrections~\cite{Gronau:1995hm}.
While the data can be used to constrain some of these corrections, completely relaxing the \SUF assumption is not possible.
We proceed by including \textit{factorizable} \SUF-breaking corrections motivated by the assumed factorization of the different decay topologies.
These corrections then enter through form factors, decay constants and meson masses and are at the level of $20$--$30\%$.
We find that, including factorizable \SUF breaking, we can perfectly accommodate the available data.\\

This work is outlined as follows: we introduce our notation, observables and inputs in Sec.~\ref{sec:observables}.
We then discuss the full \SUF analysis of the $B\to PP$ decays, which we perform using the topological parametrisation.
In Sec.~\ref{sec:factsuf}, we introduce factorizable \SUF breaking, which can be easily incorporated using the QCD factorization parametrisation.
We discuss our results in detail in Sec.~\ref{sec:SU3analysis} and make posterior predictions for all observables, including unmeasured (mixing-induced) CP asymmetries.
Several decay modes that would give additional information are highlighted.
We also discuss the obtained fit parameters and compare these briefly with QCD factorization results, leaving a more in-depth discussion for future work.
We conclude in Sec.~\ref{sec:conclusion}. 

\section{Preliminaries}

\subsection{Observables} \label{sec:observables}
We focus on $\bar{B}_q \to M_1M_2$ transitions, considering only light pseudoscalar mesons $M_{1,2} = \pi, K$.
These decays are mediated by $b \to p$ transitions, with $p = d, s$ depending on the initial and final state mesons.
In full generality, the amplitude of such decays can be written as:
\begin{equation}
    \label{eq:general_amplitude}
    \mathcal{A}(\bar{B}_q \to M_1 M_2)=i\frac{G_F}{\sqrt{2}}\left[\lambda_u^{(p)} A^{ut}_{p} +\lambda_c^{(p)} A^{ct}_{p}\right],
\end{equation}
using the CKM unitarity relation $\lambda_u^{(p)} +\lambda_c^{(p)} + \lambda_t^{(p)} = 0$, where $$\lambda_i^{(p)}\equiv V_{ib}V_{ip}^* \ .$$
The CP conjugated process $\mathcal{\bar{A}} \equiv \mathcal{A}(B\to \widebar{M_1} \widebar{M_2})$, is obtained by taking the CP conjugate of the CKM elements entering through $\lambda_i^{(p)}$, which fixes the CP sign convention for the $B_q$ meson.

The CP-averaged branching ratio is then given by 
\begin{equation}\label{eq:breq}
\mathcal{B}(\bar{B}_q \to M_1 M_2) = S \, \frac{\tau_{B_q}}{32\pi m_{B_q}} \, \Phi\left(\frac{m_{M_1}}{m_{B_q}}, \frac{m_{M_2}}{m_{B_q}}\right)
\left({|\mathcal{A}|^2 + |\mathcal{\bar{A}}|^2}\right) \ ,    
\end{equation}
where $S=1/2$ if $M_1 = M_2$ and $S=1$ otherwise. The phase space function $\Phi(x,y)$ is given by:
\begin{equation}
\Phi(x,y)=\sqrt{\left[1-(x+y)^2\right]\left[1-(x-y)^2\right]}   \ .  
\end{equation}
We note that this accounts for \SUF-breaking effects in the phase space. Numerically, however, the phase space factor only gives breaking effects smaller than $2\%$.

The direct CP asymmetry is defined as \begin{equation}\label{eq:adir}
\mathcal{A}_\text{CP}^\text{dir}(B^+\to f) = \frac{|\mathcal{A}(B^+\to f)|^2 - |\mathcal{A}(B^-\to \bar f)|^2}{|\mathcal{A}(B^+\to f)|^2 + |\mathcal{A}(B^-\to \bar f)|^2} \ ,
\end{equation}
which holds for $B^+$ decays and neutral $B$ mesons decaying to flavour-specific final states. 

Neutral $B_q^0$ mesons decaying to a CP eigenstate provide additional information due to $B_q^0-\bar B_q^0$ mixing effects, experimentally probed through time-dependent analyses.
We define the time-dependent CP asymmetry as \cite{Fleischer:2002ys}
\begin{equation}
\mathcal{A}_\text{CP}(t)=\frac{\Gamma (B^0_q(t)\to f) - \Gamma (\bar{B}^0_q (t) \to f)}{\Gamma (B^0_q(t)\to f)+\Gamma (\bar{B}^0_q(t)\to f)}=\frac{\mathcal{A}_\text{CP}^\text{dir}\cos{(\Delta M_q t)} + \mathcal{A}^\text{mix}_\text{CP}\sin{(\Delta M_q t)}}{\cosh{(\Delta \Gamma_q t/2)} + \mathcal{A}^{\Delta \Gamma}_\text{CP}\sinh{(\Delta \Gamma_q t/2)}},
\end{equation}
where $\Delta M_q \equiv M_H^{(q)} - M_L^{(q)}$ and $\Delta \Gamma_q \equiv \Gamma_L^{(q)} - \Gamma_H^{(q)}$ are the mass and decay-width differences between the heavy and light mass eigenstates of the $B^0_q$ system. The direct and mixing-induced CP asymmetries are defined as
\begin{equation}
    \mathcal{A}_\text{CP}^\text{dir} \equiv \frac{1-|\xi_f|^2}{1+|\xi_f|^2} \,, \qquad
    \mathcal{A}^\text{mix}_\text{CP} \equiv \frac{2\ \text{Im}(\xi_f)}{1+|\xi_f|^2} ,
\end{equation}
in terms of the convention-independent parameter
\begin{equation}
    \xi_f= - e^{-i\phi_q}\frac{\mathcal{A}(\bar{B}^0_q\to f)}{\mathcal{A}(B^0_q\to f)}.
\end{equation}
In the full process, the $B_q^0$--$\bar{B}_q^0$ mixing phase $\phi_q = 2\;{\rm Arg}(V_{tq}^* V_{tb})$, which is convention dependent, combines with the phases of the $B_q^0\to f$ decay.
In our analyses, we include CP violation in the decays but not in the mixing of the $B$ and $K$ mesons\footnote{
    Mass eigenstates are defined in terms of flavour eigenstates as: $p|B^0_q\rangle+q|\bar B_q^0\rangle$.
    CP violation in the mixing would imply $|q/p| \neq 1$ and consequently $\xi_f$ should be corrected since $\xi_f ~\propto~ q/p$.
}.

For $B_d^0$ mesons, the relative width difference $\Delta\Gamma_d/\Gamma_d \sim \mathcal{O}(10^{-3})$ is negligible. However, $B_s^0$ mesons have $y_s \equiv \Delta \Gamma_s/(2\Gamma_s) = 0.0635$~\cite{PDG2024}, which provides access to an additional CP observable  
\begin{equation}\label{eq:AdelGam}
\mathcal{A}_\text{CP}^{\Delta \Gamma}\equiv \frac{2\,\text{Re} (\xi_f)}{1+|\xi_f|^2} \ .
\end{equation}
Finally, the three CP observables are related by the unitarity condition
\begin{equation}\label{eq:SR_CP}
    \big(\mathcal{A}_\text{CP}^\text{dir}\big)^2
  + \big(\mathcal{A}_\text{CP}^\text{mix}\big)^2
  + \big(\mathcal{A}_\text{CP}^{\Delta \Gamma}\big)^2 = 1 \ .
\end{equation}

The significant width difference in the $B_s^0$ system introduces a subtle complication for branching ratio determinations, which are typically obtained from time-integrated untagged rates.
However, the theoretical expressions are calculated at $t=0$, introducing a difference between the calculated and measured branching ratios.
Correcting for this effect gives
\cite{DeBruyn:2012wj}
\begin{equation}\label{eq:breqtheo}
    \mathcal{B}(B_s^0\to f)_\text{exp} = \mathcal{B}(B_s^0\to f)_\text{theo} \left(\frac{1+\mathcal{A}^{\Delta\Gamma}_\text{CP} \,y_s}{1-y_s^2}\right) \ .
\end{equation}
In the following, we refer to all our predicted branching ratios as $\mathcal{B}_{\rm exp}$, although the difference is only significant for $B_s^0$ decays.
For decays into flavour-specific final states, $\mathcal{A}^{\Delta\Gamma}_\text{CP} = 0$ such that the conversion factor reduces to $1-y_s^2$.


\subsection{CKM Inputs}
For the CKM matrix, we employ the Wolfenstein parametrisation and its parameters extracted from a global unitarity fit of the CKM matrix.
We use the results presented in Ref.~\cite{Charles:2004jd}, which are consistent with those of Ref.~\cite{UTfit:2022hsi} within the quoted uncertainties.
Assuming symmetric Gaussian distributions, the Wolfenstein fit parameters read
\begin{align}\label{eq:ckm}
    A          &= 0.81975 \pm 0.00645 \ ,  &\lambda&    = 0.22499 \pm 0.00022 \ , \\
    \bar{\rho} &= 0.1598  \pm 0.0076 \ ,  &\bar{\eta}& = 0.3548 \pm 0.0054 \ ,
\end{align}
which gives $\gamma = (65.75 \pm 1.07)^\circ$, in agreement with the latest LHCb average $\gamma|_\mathrm{LHCb} = (64.6 \pm 2.8)^\circ$~\cite{Hao:2024ggu,LHCb-CONF-2024-004}.

For $\lambda^{(p)}_i = V_{ib}V_{ip}^*$, we then find
\begin{align}
    \lambda_u^{(d)} &= (1.49 - 3.31 i) \times 10^{-3} \ , {}& \lambda_c^{(d)}&= (-9.33 + 0.0057 i) \times 10^{-3} \ , \\
    \lambda_u^{(s)} &= (3.44 - 7.65 i) \times 10^{-4} \ , &  \lambda_c^{(s)}&= (4.04 + 0.00013 i) \times 10^{-2} \ .
\end{align}

For the $B_q^0$--$\bar{B}_q^0$ mixing phase we find, in the SM,
\begin{equation}
    \phi_d = (45.68^{+0.66}_{-0.60})^\circ, \qquad
    \phi_s = -0.03764 ^{+0.0052}_{-0.0056} .
\end{equation}
We note that the value of $\phi_d$ agrees within uncertainties with the analysis of $B_d^0\to J/\psi K_S^0$ decays including penguin pollution \cite{Barel:2022wfr, Barel:2020jvf}.

\subsection{Experimental inputs}\label{sec:expinputs}
The experimental data for the branching ratios is given in~\cref{tab:table_topofit_BR}, which we take from the PDG \cite{PDG2024} unless otherwise specified.
In general, these results are averages of several experimental measurements, and the experimental correlations are not quoted or known.

Six decay modes are measured as ratios with respect to $B\to K\pi$ decays and are given in~\cref{tab:table_topofit_ratBR}. 
For easy comparison, we quote the values of the corresponding branching ratios in~\cref{tab:table_topofit_BR}, using the PDG average for the normalization channel.
However, in our numerical analysis, we directly include the measured ratios as inputs.
If a decay is measured directly and through a normalization channel, both are included in our analysis. 

The $B_s^0$ measurements from the LHCb collaboration \cite{LHCb:2012ihl, LHCb:2016inp} depend on the ratio of $B_s^0$ \textit{vs} $B^0/B^+$ mesons production fraction, $f_s/f_d$.
We adopt $f_s/f_d = 0.239$~\cite{LHCb:2021qbv}, which is obtained at a collision energy of 7 TeV.
Although fragmentation fractions depend on the collision energy, we have verified that this dependence does not impact our results for the current experimental sensitivity. 
Finally, we note that the $B_s^0\to K^0\bar{K}^0$ channel was also measured normalized to the $B^0\to K^0 \phi$ mode, for which we use the PDG average \cite{PDG2024}.

The measured direct CP asymmetries are given in~\cref{tab:table_topofit_Adir}, and the mixing-induced CP asymmetries are listed in~\cref{tab:table_topofit_Amix}.
For the $B^+ \to K^+ \bar{K}^0$ and $B^0\to K^0 \bar{K}^0$ decays, the listed direct CP asymmetries are obtained from averaging the PDG results for the $K^0$ modes with those obtained in Refs.~\cite{LHCb:2013vip} and \cite{Belle:2006uzg} for the $K_S^0$ modes, with $K_S^0=\frac{1}{\sqrt{2}}\left(K^0-\bar K^0\right)$.
We quote results for the $K^0$ and $\bar{K}^0$ modes, which is relevant for the branching ratios, except for the mixing-induced CP asymmetries where, for clarity, we report decays to $K_S^0$.

The CP asymmetries for $B_s^0 \to K^+ K^-$ have been measured exclusively by the LHCb collaboration \cite{LHCb:2020byh}, and correlations between the CP asymmetries are provided. 
Consequently, we incorporate these correlations for this channel in our analysis. 
In addition, the LHCb collaboration measured $\mathcal{A}^{\Delta\Gamma}_{\rm CP}$ \cite{LHCb:2020byh}:
\begin{equation}
\mathcal{A}^{\Delta\Gamma}_{\rm CP}(B_s^0\to K^+K^-) = -0.897\pm 0.087  \ .
\end{equation}
As this observable was obtained without assuming the unitarity constraint in \cref{eq:SR_CP}, we also include this as an independent observable in our fit, accounting for correlations. 

Finally, we note that the difference between the theoretical and experimental branching ratios, determined through \eqref{eq:breqtheo}, results in a $\mathcal{O}(5\%)$ effect. This highlights the importance of incorporating this correction, as also discussed in Ref.~\cite{Fleischer:2016ofb}.

For the masses and lifetimes, we use \EOS's default parameters~\cite{EOS:v1.0.14}.

\section{\boldmath \SUF analysis of $B\to PP$}
\subsection{Parametrisation of the amplitudes} \label{sec:SUF_parametrisation}
Typically, \SUF analyses are set up by decomposing the $B\to PP$ decays using a topological parametrisation, \SUF irreducible representations \cite{He:2018php, He:2018joe, Fu:2002nr, Huber:2021cgk} or reduced matrix elements~\cite{Berthiaume:2023kmp, Bhattacharya:2025wcq}.
In the \SUF-limit, all these approaches were shown to be equivalent~\cite{He:2018php, Shi:2025eyp}.
In the following, we work with the topological parametrisation. 

To obtain the amplitudes for $B \to PP$ decays in terms of the topological diagram amplitudes, we write the pseudoscalar meson matrix $M_j^i$ as:
\begin{equation}
\label{eq:Mmatrix}
\begin{aligned}
   M &= \left( \begin{array}{ccc} 
    \frac{\pi^0}{\sqrt 2} + \frac{\eta_8}{\sqrt 6}
     & \pi^- & K^- \\
    \pi^+ & - \frac{\pi^0}{\sqrt2} + \frac{\eta_8}{\sqrt 6}
      & \bar K^0 \\
    K^+ & K^0 & -2 \frac{\eta_8}{\sqrt 6}  \end{array} \right)
\end{aligned}
+ 
\begin{aligned}
   \left( \begin{array}{ccc} 
   \frac{\eta_0}{\sqrt 3}
     & 0 & 0 \\
    0 & \frac{\eta_0}{\sqrt 3}
      & 0 \\
    0 & 0 & \frac{\eta_0}{\sqrt 3}  \end{array} \right),
\end{aligned}
\end{equation}
where the first term corresponds to the \SUF flavour meson octet.
The second term represents the singlet state $\eta^0$, described by independent singlet topologies. 
For completeness, we include these terms in our parametrisation below, but we do not include these modes in our analysis as discussed in \cref{sec:SU3analysis:fitsetup}.
The $B$-meson vector is given by: $B_i = (B^+, B^0, B^0_s) = \left( B(\bar{b}u), B(\bar{b}d), B(\bar{b}s) \right)$.

We parametrise the amplitude as in \eqref{eq:general_amplitude}, where $A^{ut}$ represent tree-like topologies and $A^{ct}$ are the penguin-like topologies. 
Within the topological parametrisation, we then have\footnote{We adopt the same notation for the topologies as Ref.~\cite{He:2018php}, to which we also refer for details on the electroweak penguin parameters and their definitions.}
\begin{align}
A^{ut}_{p,\rm topo} \equiv & \quad  T~B_i (M)^{i}_j  \bar H^{jl}_k  (M)^k_l   + C~B_i (M)^{i}_j \bar H^{lj}_k  (M)^k_l
+ A~B_i  \bar H^{il}_j   (M)^j_k (M)^{k}_l\nonumber \\
&+ E~B_i \bar H^{li}_j (M)^j_k (M)^{k}_l
+ T_{ES}~ B_i  \bar H^{ij}_{l}   (M)^{l}_j    (M)^k_k
+T_{AS}~ B_i \bar H^{ji}_{l}  (M)^{l}_j  
(M)^k_k\nonumber\\
&+ T_{S}~ B_i  (M)^{i}_j   \bar H^{lj}_{l}  (M)^k_k + T_{PA}~ B_i \bar H^{li}_{l}  (M)^j_k (M)^{k}_j +T_{P}~ B_i (M)^{i}_j   (M)^j_k \bar H^{lk}_{l}\nonumber\\ 
&    + T_{SS}~ B_i  \bar H^{li}_{l}  (M)^j_j (M)^{k}_{k},
\label{eq:Topo_tree}
\end{align}
with $\bar{H}_1^{12} = \delta_{pd}$,  $\bar{H}_1^{13} = \delta_{ps}$ and all the other components of $\bar{H}$ are zero. Here $T_S, T_{AS}, T_{ES}$ and $T_{SS}$ are the singlet parameters.

The penguin amplitude $A^{ct}_{p,\rm topo}$ is obtained using the same formula and applying the following substitution between tree and penguin coefficients: 
\begin{align}
T&\to P_{T}, & C & \to P_{C}, &
A&\to P_{TA},& T_{P}&\to P, &
E&\to P_{TE},\nonumber\\
T_{PA}&\to P_{A}, &
T_{AS}&\to P_{AS},&
T_{ES}&\to P_{ES}, &
T_{SS}&\to P_{SS}, &
T_{S}&\to S.
\label{eq: TPrelationTop}
\end{align}

The amplitudes contributing to a given decay $B\to MM$ are obtained using \cref{eq:Topo_tree,eq: TPrelationTop} and symmetrizing over the final state mesons $A^{ut}_{q,\text{topo}}(\bar{B}\to MM) = A^{ut}_{q,\text{topo}}(\bar{B}\to M_1M_2) + A^{ut}_{q,\text{topo}}(\bar{B}\to M_2M_1)$.
Applying this procedure, we agree with the coefficients presented in Refs.~\cite{He:2018php,Huber:2021cgk}.
In total, there are 10 tree-like and 10 penguin-like complex parameters.
By comparing to the irreducible \SUF parametrisation, one can show that only 9 of each of these parameters are independent~\cite{He:2018php}.

A further reduction to 7 parameters is possible by assuming EWP-tree relations~\cite{Neubert:1998jq, Neubert:1998pt, Gronau:1995hn}, which link the penguin coefficients ($P_T, P_C, P_{TA}, P_{TE}$) to the tree parameters and ratios of the $C_{9,10}$ to $C_{1,2}$ Wilson coefficients (see also the recent discussion in Refs.~\cite{Shi:2025eyp, Bhattacharya:2025wcq}).
These relations require neglecting the contributions of the $\mathcal{O}_8$ and $\mathcal{O}_7$ operators. In the SM, the Wilson coefficients of these operators are very small compared to $C_{9,10}$ and, as such, often neglected. In our default analysis, we do not impose EWP-tree relations in order to remain independent of any (SM) assumptions. 

Finally, we also note that even when neglecting $\mathcal{O}_8$ and $\mathcal{O}_7$, the EWP-tree relations assume that penguin contractions of current-current operators $\mathcal{O}_{1,2}$ can be neglected. 

\subsection{Sum rules and relations between amplitudes}\label{sec:su3_SR}
Before turning to a full \SUF analysis of the available data, it is interesting to look at subsystems of the $B\to PP$ decays.
Often considered are isospin tests of the $B\to \pi\pi$ and $B\to \pi K$ subsystems \cite{Gronau:1990ka,Gronau:1995hn,Fleischer:2008wb,Fleischer:2018bld},
which are usually limited by the experimental precision of the (CP asymmetries of the) modes with neutral pions.
Isospin sum rules at the amplitude level can be read directly from Table~\ref{tab:tableAmplBPPQCDF}.
Similarly, using the decay topologies, we can quickly identify $U$-spin partners between the $b\to s$ and $b\to d$ transitions, namely decays that have identical decay topologies and only differ by their respective CKM factors.

For neutral $B_q^0$ decays, which have both the direct and mixing-induced CP asymmetries, the experimental data could be used directly to extract the penguin-tree ratio of the $b\to p$ transition\footnote{
Our notation is related to Ref.~\cite{Fleischer:2016ofb} by $\theta^{(\prime)} \to \theta_{d(s)}$ and $d^{(\prime)} \to r_{d(s)}/R_b$, where $R_b = (1-\lambda^2/2)\frac{1}{\lambda} \left|\frac{V_{ub}}{V_{cb}}\right| \simeq 0.39$ measures a side of the Unitarity Triangle.
}.
\begin{equation}\label{eq:rat}
    r_p e^{i\theta_p} \equiv \frac{A^{ct}_{p,\rm topo} }{A^{ut}_{p,\rm topo} } \ ,
\end{equation}
where $\theta_p$ is a strong phase, $p = d, s$ and $(r_p,\theta_p)$ are decay specific. For each decay, $A_{q,\rm topo}^{ut}$ can be obtained through \eqref{eq:Topo_tree} and equivalently for the penguin amplitude $A_{q,\rm topo}^{ct}$.

\begin{figure}[t]
    \centering
    \includegraphics[width=0.8\textwidth]{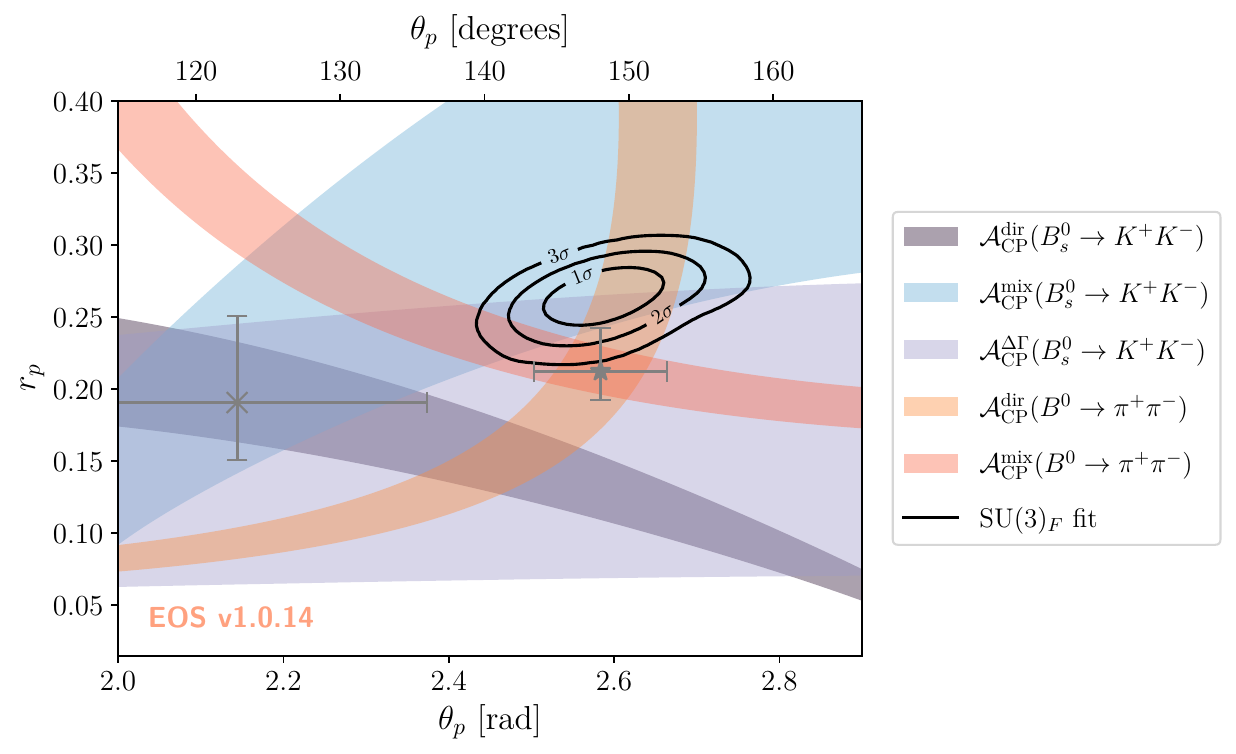}
    \caption{
        $68\%$ probability intervals of the (partially correlated) experimental constraints on the $B_d^0\to\pi^+\pi^-$ and $B_s^0\to K^+K^-$
        CP observables in the $(r_p, \theta_p)$ plane defined in \cref{sec:su3_SR}.
        The $1, 2$ and $3 \sigma$ postdictions of the global \SUF fit are overlaid; see \cref{sec:SU3analysis} for details.
        The grey star and cross show the best-fit points from the CP asymmetries of $B_d^0\to\pi^+\pi^-$ and $B_s^0\to K^+K^-$, respectively.
    }
    \label{fig:dtheta}
\end{figure}

The only $U$-spin partner decays for which both CP asymmetries have been measured are $B^0_d\to\pi^+\pi^-$ and $B^0_s\to K^+K^-$, mediated via a $b\to d$ and a $b\to s$ transition, respectively.
These decays have already been studied in several analyses \cite{Fleischer:1999pa,Fleischer:2007hj,Fleischer:2010ib,Fleischer:2016ofb,Fleischer:2018bld, Fleischer:2022rkm, Nir:2022bbh}, as they can be used to extract the CKM angle $\gamma$ and $\phi_s$ modulo small \SUF breaking effects \cite{Fleischer:2016jbf, Fleischer:2022rkm}.
Using the experimental measurements for the CP asymmetries given in~\cref{tab:table_topofit_Adir,tab:table_topofit_Amix,tab:table_topofit_ADG}, we obtain constraints on $(r_d,\theta_d)$ and $(r_s,\theta_s)$ from $B^0_d\to\pi^+\pi^-$ and $B^0_s\to K^+K^-$, respectively. 
The $1\sigma$ uncertainty constraints on the CP asymmetries of these two decays are shown in \cref{fig:dtheta}, together with the best-fit point for $(r_d,\theta_d)$ and $(r_s,\theta_s)$ \footnote{%
    For $B_s^0\to K^+K^-$, we also use $\mathcal{A}_{\rm CP}^{\Delta\Gamma}$.
    Although this observable is related to the direct and mixing-induced CP asymmetries via \cref{eq:SR_CP}, this constraint was not used in the experimental analysis~\cite{LHCb:2020byh}.
    This measurement provides, therefore, additional information on the fit.
}.
We obtain
\begin{equation}\label{eq:rds}
    (r_d, \theta_d) = (0.21^{+0.03}_{-0.02}, 2.58^{+0.08}_{-0.08}),
    \qquad
    (r_s, \theta_s) = (0.19^{+0.06}_{-0.04}, 2.14^{+0.23}_{-0.24}).
\end{equation}
These findings agree with the recent analysis of Ref.~\cite{Fleischer:2022rkm}.
In the $U$-spin limit, we have $r = r_d = r_s$ and $\theta = \theta_d = \theta_s$.
Comparing these numerical values, we find $20-30\%$ differences due to \SUF breaking in this subset of decays, which is in agreement with the analyses in~\cite{Fleischer:2022rkm, Nir:2022bbh}.

Fitting $r$ and $\theta$ to the experimental values yields a minimal $\chi^2$ of 8.13 for 3 degrees of freedom.
We conclude, therefore, that already in the $\left(B^0_d\to\pi^+\pi^-, B^0_s\to K^+K^-\right)$ system, \SUF symmetry is violated at the $2\sigma$ level. 

The $U$-spin partners $\left(B^0_d\to K^+\pi^-, B^0_s\to\pi^+K^-\right)$, recently studied in detail in Refs.~\cite{Fleischer:2018bld, Fleischer:2022rkm}, are flavour-specific and exhibit no mixing-induced CP asymmetry.
The simple analysis done above, based on only the CP asymmetries, can, therefore, not be performed. 
For other $U$-spin systems, notably the $(B_s^0\to \pi^+\pi^-, B_d^0\to K^+K^-)$ which would provide information on the exchange and penguin annihilation topologies, there is not yet sufficient information available on the CP-asymmetries to perform the above analysis.
We refer to e.g.~\cite{Fleischer:2016ofb} for a discussion on these modes. 

\begin{figure}[t]
    \centering
    \includegraphics[width=0.75\textwidth]{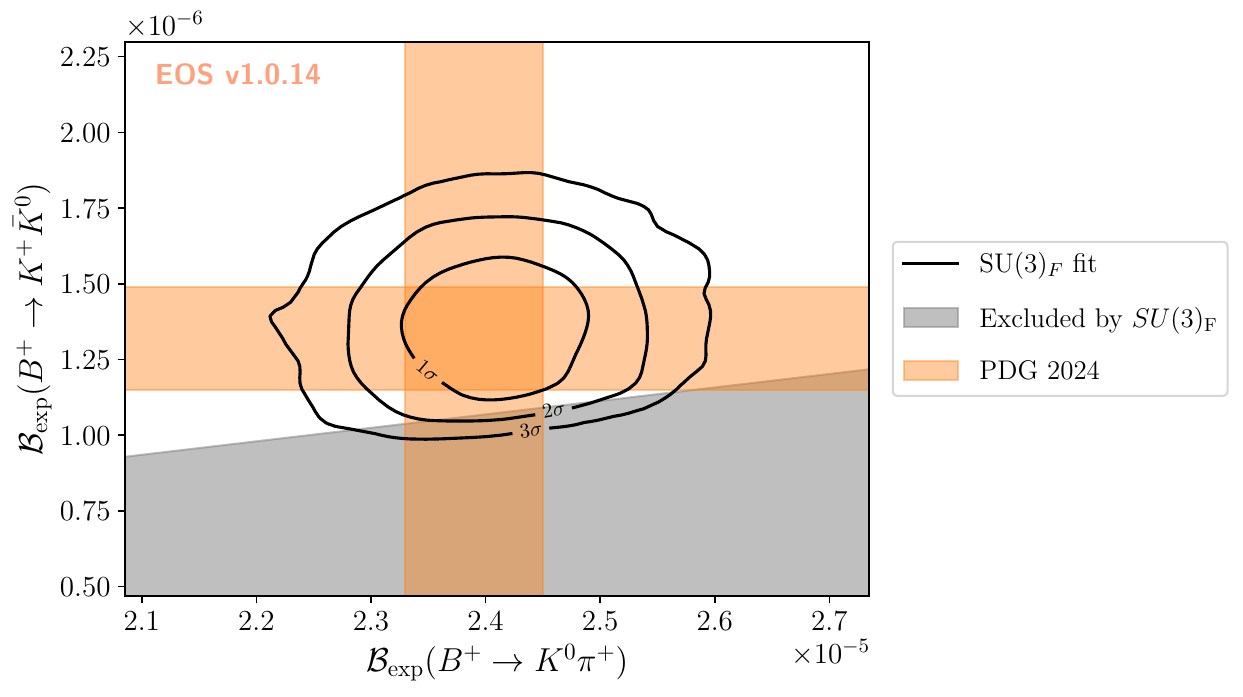}
    \caption{
        $68\%$ probability intervals of the experimental constraints on the $B^+\to K^+\bar{K}^0$ and $B^+\to K^0\pi^+$ branching ratios.
        The $1, 2$ and $3 \sigma$ postdictions of the global \SUF fit are overlaid; see \cref{sec:SU3analysis} for details.
        The \SUF symmetry excludes the grey region.
    }
    \label{fig:BR_relation}
\end{figure}

In addition, \SUF symmetry also puts constraints on the branching ratios of $U$-spin partners.
Using \cref{eq:general_amplitude}, and $\lambda_c = -\lambda_t - \lambda_u$, we obtain
\begin{equation}
    \frac{\mathcal{B}^{b\to d}}{\mathcal{B}^{b\to s}} =
    \left| \frac{\lambda_u^{(d)} A^{ut}_d + \lambda_c^{(d)} A^{ct}_d}{\lambda_u^{(s)} A^{ut}_s + \lambda_c^{(s)} A^{ct}_s} \right|^2 =
    \left| \frac{\lambda_t^{(d)}}{\lambda_t^{(s)}} \right|^2
    \left| \frac{1 + \frac{\lambda_u^{(d)}}{\lambda_t^{(d)}} \frac{A^{ct}_d - A^{ut}_d}{A^{ct}_d}}{1 + \frac{\lambda_u^{(s)}}{\lambda_t^{(s)}} \frac{A^{ct}_s - A^{ut}_s}{A^{ct}_s}} \right|^2.
\end{equation}
Using that, numerically, we have $\mathrm{Im}\frac{\lambda_u^{(d)}}{\lambda_t^{(d)}} \gg \mathrm{Re}\frac{\lambda_u^{(d)}}{\lambda_t^{(d)}}, \mathrm{Re}\frac{\lambda_u^{(s)}}{\lambda_t^{(s)}}, \mathrm{Im}\frac{\lambda_u^{(s)}}{\lambda_t^{(s)}}$, and that $U$-spin partners satisfy $A^{ut}_{s} = A^{ut}_{d}$ and $A^{ct}_{s} = A^{ct}_{d}$, we get
\begin{equation}
    \frac{\mathcal{B}^{b\to d}}{\mathcal{B}^{b\to s}} \simeq
    \left| \frac{\lambda_t^{(d)}}{\lambda_t^{(s)}} \right|^2
    \left( 1 + \left| \frac{\lambda_u^{(d)}}{\lambda_t^{(d)}} \frac{A^{ct}_d - A^{ut}_d}{A^{ut}_d} \right|^2 \right) >
    \left| \frac{\lambda_t^{(d)}}{\lambda_t^{(s)}} \right|^2 \simeq 0.042.
\end{equation}
This constraint is, \eg, relevant for the $\left(B^+\to K^+\bar{K}^0, B^+\to K^0\pi^+\right)$ system where it excludes $10\%$ of the experimentally allowed region, as depicted in \cref{fig:BR_relation}. 

Considering these subsystems provides insight into the validity of the $SU(3)_F$ assumption.
The analysis of the $\left(B^0_d\to\pi^+\pi^-, B^0_s\to K^+K^-\right)$ subsystem already shows that the data cannot be described by a parametrisation that respects this symmetry.
Of course, the key point of a parametrisation based on $SU(3)_F$ is that it links all the $B\to hh$ decays; a full study of its validity requires a complete analysis of all the modes.
We perform this analysis in the following section.

\subsection{Full \SUF analysis of $B\to PP$ decays}\label{sec:SU3analysis}
\subsubsection{Fit setup}\label{sec:SU3analysis:fitsetup}
We perform a Bayesian analysis of the real and imaginary parts of the topological coefficients.
We include all available experimental data on branching ratios, direct and mixing-induced CP asymmetries and ratios of branching ratios.
We account for CKM uncertainties by varying the Wolfenstein parameters as described in \cref{eq:ckm}.

In our nominal fit, we do not include decays into $\eta$ and $\eta'$ final states.
The treatment of these states within an \SUF analysis, with or without breaking terms, is subtle.
The singlet $\eta_0$ and octet $\eta_8$ states mix under \SUF breaking into the physical $\eta, \eta'$ states.
Within the mixing-angle description of these states \cite{Feldmann:1998su}, the \SUF limit implies that the mass eigenstates coincide with the \SUF eigenstates, and we have $|\eta\rangle = |\eta_8\rangle$ and $|\eta'\rangle = |\eta_0\rangle$ thereby decoupling the singlet parameters completely from the octet parameters. Moreover, within an \SUF analysis (even when including breaking effects), $B$ decays to $\eta'$ final states should be considered separately from the other light-meson final states.
We refer to Ref.~\cite{Bolognani:2024zno} for a recent discussion on treating $\eta$ and $\eta'$ decay modes in nonleptonic $D$ decays.
Since there is currently limited experimental data for the $B\to P\eta'$ decays, we postpone such an analysis to future work. 

We also do not include decays to $\eta$ final states. As argued above, the $B_d^0\to \pi^+\pi^-$ and $B_s^0\to K^+K^-$ data already shows a deviation from \SUF symmetry at the $2\sigma$ level by considering only these decays.
In the following, we want to explore further if \SUF symmetry can describe the data for the light-meson final state and, specifically, which experimental data drives the determination of the parameters before enlarging the analysis to the $\eta$ final states.
However, we do postdict branching ratios and CP asymmetries for $B$ to $\eta$ decays following our \SUF analysis.\\
Excluding the singlet parameters, 6 complex tree and 6 complex penguin parameters remain.
As mentioned above, two of these parameters are redundant.
This leaves a choice on which parameter to eliminate to be consistent with the \SUF irreducible representations \cite{He:2018php}.
Here we set $A = E$ and $P_{TA} = P_{TE}$.
Additionally, to account for the invariance under a global phase shift, we fix $\text{Arg}~T=0$, leaving 19 real parameters.

Our predictions for the values of the parameters, $\vec{\theta}$, are made through the maximization of the posterior Probability Density Function (PDF), $P(\text{experimental data}\,|\,\vec{\theta})$.
This PDF is estimated by sampling the parameter space assuming uniform coefficient priors.
The support of these priors is purely data-driven, \ie no theoretical inputs have been taken into account for the possible values of the coefficients.

To evaluate the quality of the fit, we calculate the global $\chi^2$ value at the best-fit point, 
\begin{equation}
    \chi^2=-2\ln P(\text{experimental data}\,|\,\vec{\theta})
\end{equation}
as well as the corresponding \pvalue, which measures the goodness of fit between our theoretical model and the experimental data. 

This analysis is performed using \EOS~\cite{EOSAuthors:2021xpv} version v1.0.14~\cite{EOS:v1.0.14}, a publicly available software specifically designed for studies in flavour physics phenomenology.
We draw all posterior samples using the nested sampling algorithm~\cite{Higson:2018} as implemented in the \texttt{dynesty} software~\cite{Speagle:2020,dynesty:v2.0.3}.
The codes used to run our analysis and all our results are available in the analysis repository~\cite{EOS-DATA-2025-01}.

\subsubsection{Results}\label{sec:SU3analysis:results}
At the best-fit point, we obtain $\chi^2 = 32.3$ for 15 degrees of freedom (34 constraints for 19 parameters), yielding a \pvalue of $0.58\%$.
This small \pvalue is below our \textit{a-priori} threshold of 3\%, and we conclude that this fit is not satisfactory.

As anticipated in \cref{sec:su3_SR}, the main tensions in the fit are due to the $\left(B^0_d\to\pi^+\pi^-\right.$, $\left.B^0_s\to K^+K^-\right)$ system which shows a manifest tension with the \SUF symmetry, as visible in \cref{fig:dtheta}.
The recent update of these modes, driven by the measurements of the $B_s^0$ modes by the LHCb collaboration \cite{LHCb:2020byh}, already gives a very constrained picture.
This is important to keep in mind when comparing our analysis to previous \SUF analyses based on older data sets.

We observe that the posterior distributions of the topological parameters are multi-modal, highly non-Gaussian and show strong correlations between different parameters.
We also find that the tensions in the fit bias the estimation of the topology parameters, which do not follow the expected hierarchy pattern $T \sim \frac{1}{3}C > P > P_T > A, E$ from kinematic arguments. 

Despite the poor fit quality, we provide postdictions for all the $B\to PP$ observables, measured or not, in \cref{tab:table_topofit_BR,tab:table_topofit_Amix,tab:table_topofit_Adir,tab:table_topofit_ratBR}.
We also provide our postdictions for $A^{\Delta\Gamma}_{CP}$ in \cref{tab:table_topofit_ADG} of \cref{app:ADeltaGamma}.
Although this CP asymmetry is related to the direct and mixing-induced CP asymmetries through \cref{eq:SR_CP}, our postdictions for this observable cannot be extracted from the median and uncertainty intervals of our postdictions due to their non-Gaussian distribution.
Our results show that $A^{\Delta\Gamma}_{CP}$ saturates the unitarity relation for most of the $B_s^0$ modes.
This implies that the difference between the experimental and theoretical branching ratio in \cref{eq:breqtheo} is maximal, emphasizing the importance of accounting for this effect.

\begin{figure}[!t]
    \centering
    \includegraphics[width=0.45\textwidth]{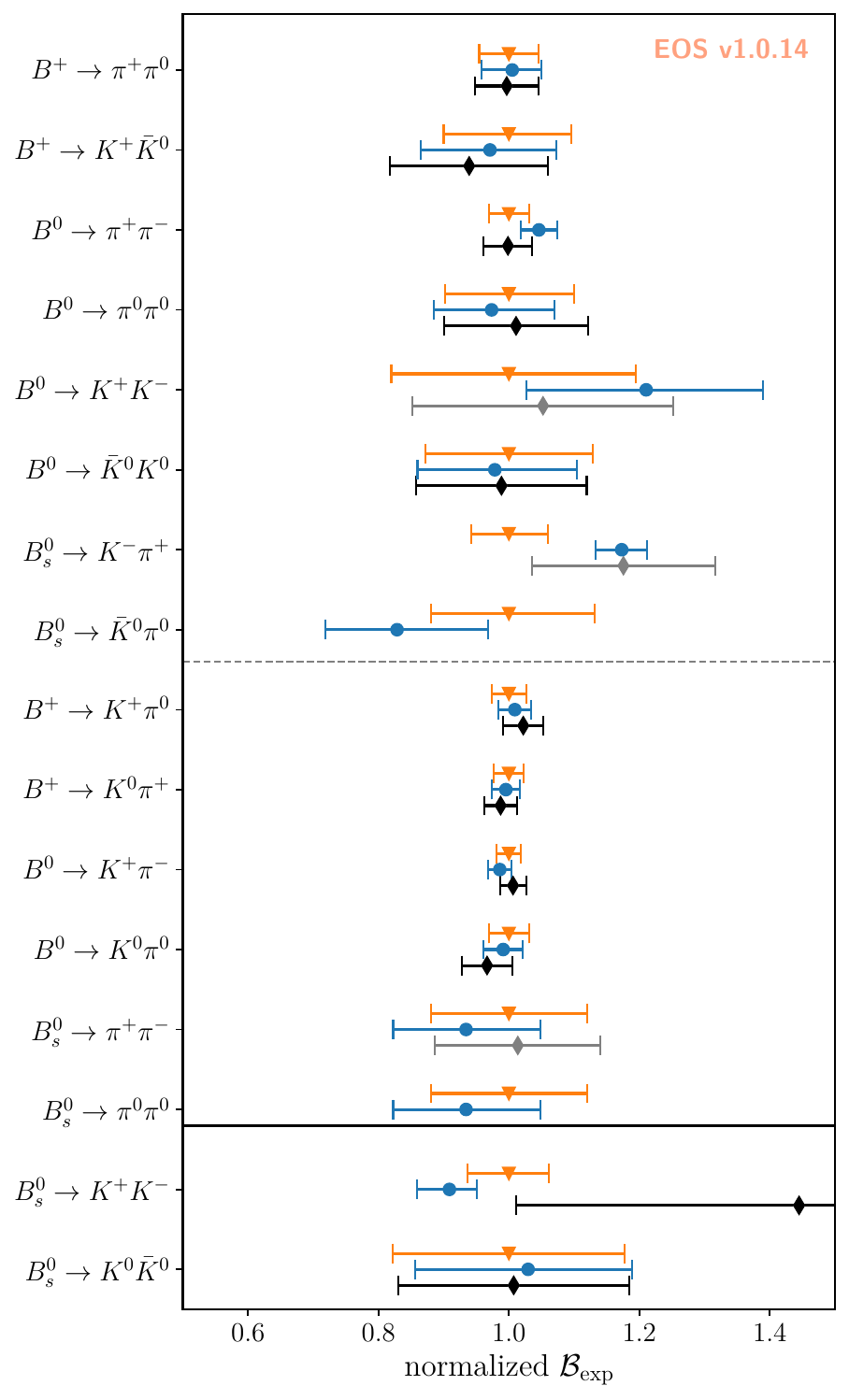} \hfill
    \includegraphics[width=0.45\textwidth]{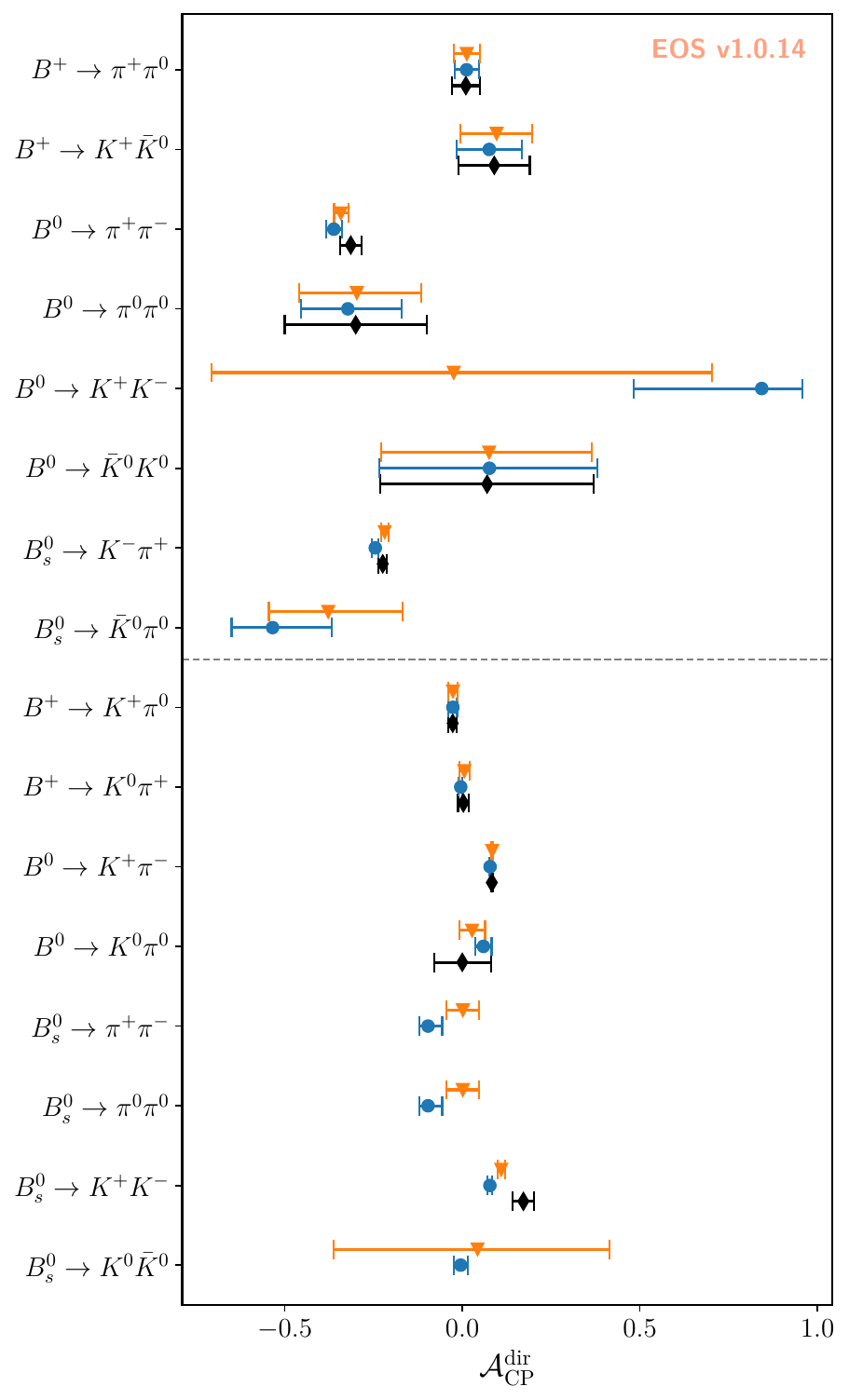} \\
    \includegraphics[width=0.45\textwidth]{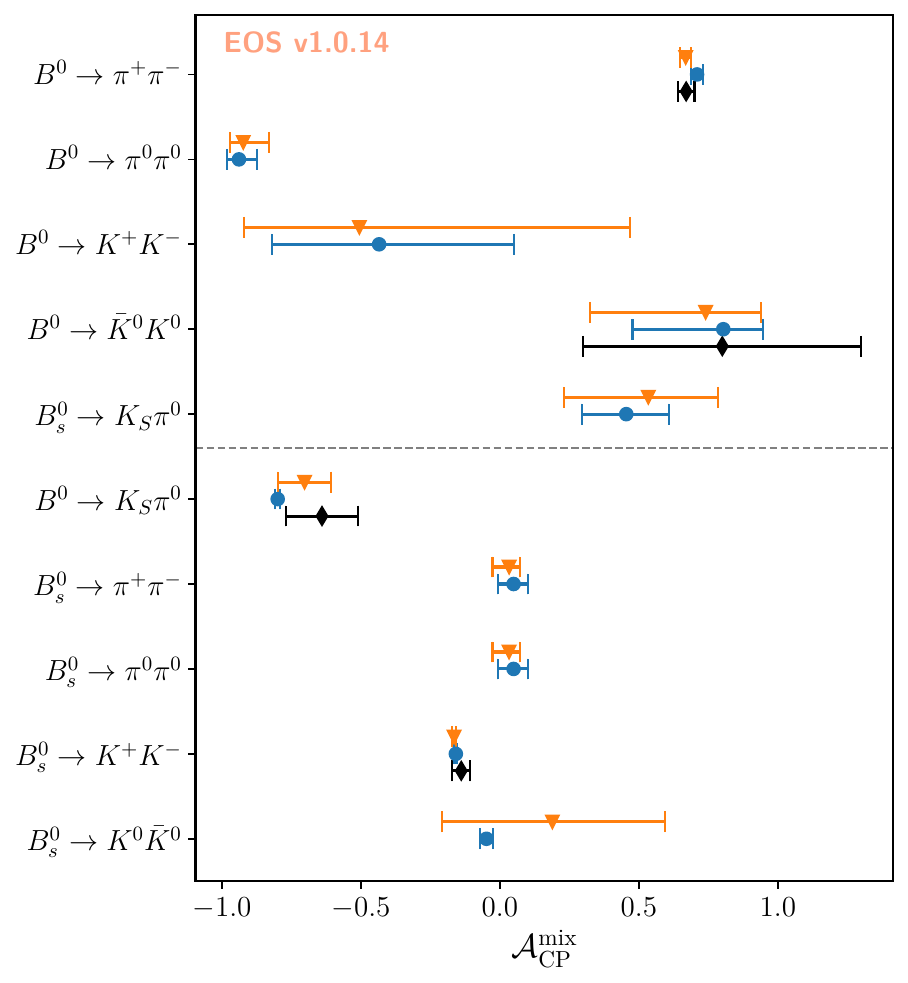} \hfill
    \includegraphics[width=0.49\textwidth]{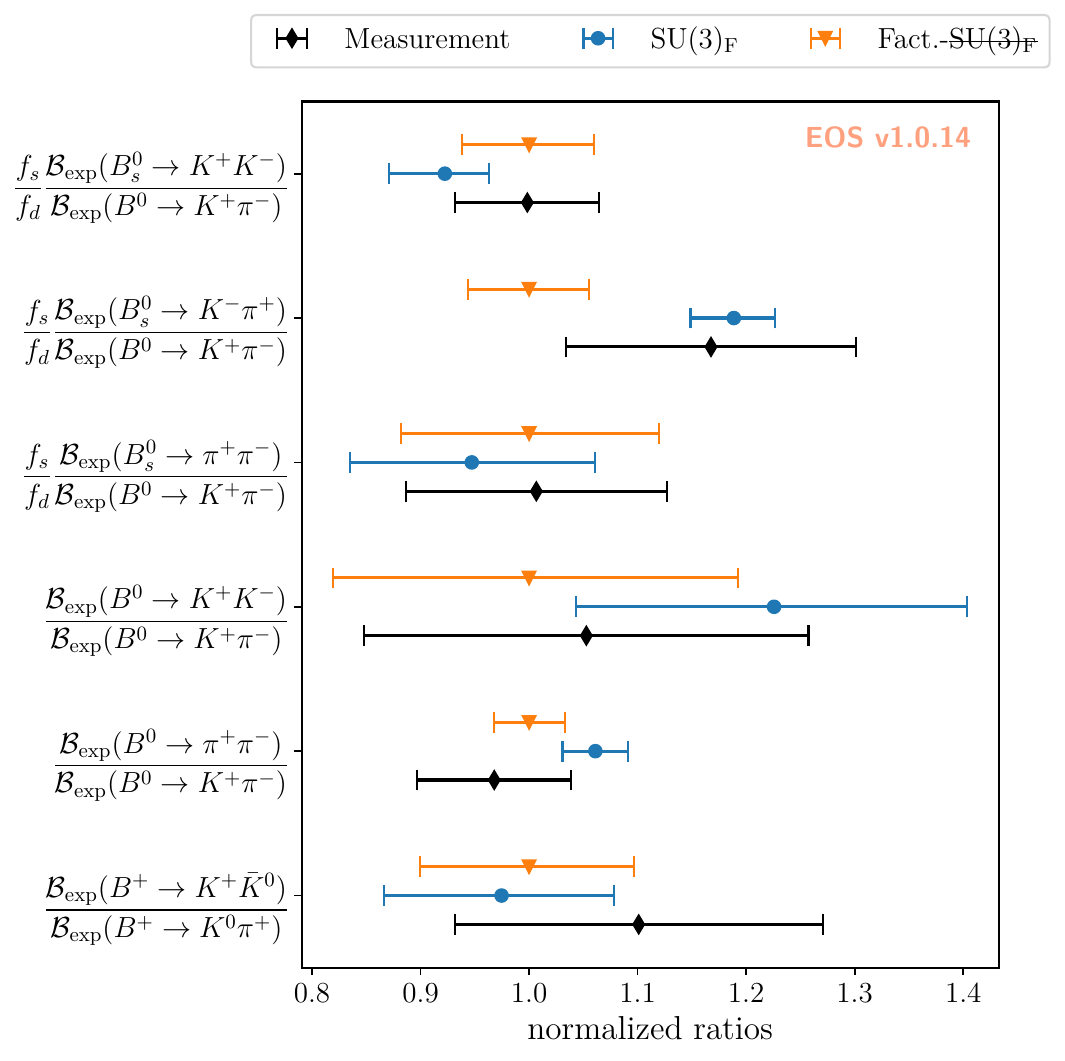}
    \caption{Predicted observables within our fit models, combined with the measured values.
        The branching ratios and their ratios are normalised to their ``Fact.-$\cancel{{\SUF}}$ '' postdictions for readability.
        Grey measurements are not directly used in the fit (ratios of branching ratios are used instead).
    }
    \label{fig:fit_noeta}
\end{figure}

Additionally, we present in \cref{fig:fit_noeta} the pull plots comparing the experimental measurements and fit postdictions under the \SUF symmetry.
The postdictions of our fit are shown in blue, experimental results used in the fit are in black, and grey is used for experimental results that are not used in the fit. 
For readability, we have normalized the branching ratios $\mathcal{B}_{\rm exp}$, defined in \cref{eq:breqtheo}, to the theory postdictions.
Most postdictions agree with the experimental results at the $1\sigma$ level, but some tensions are clearly visible in \eg $B_s^0\to K^+K^-$ branching ratio and $B^0\to\pi^+\pi^-$ CP asymmetries.
Since correlations are not apparent in the plot, some tensions contributing to the fit's poor quality are not visible.

We highlight that our analysis differs from previous works as:
\begin{itemize}
    \item We include mixing-induced CP asymmetries and, for the first time, mixing effects in the $B_s^0$ branching ratio in the analysis.
    \item We omit the $\eta^{(\prime)}$ modes as their treatment requires particular care and involves additional fit parameters.
    \item We include correlations between experimental measurements by using ratios of branching ratios, when available, instead of multiplying these ratios by world averages.
\end{itemize}
We note, though, that the poor quality of our \SUF fit seems to contrast with the work of Ref.~\cite{Huber:2021cgk}.
This may be due, in part, to the sharper picture obtained from the new experimental data.
A recent \SUF analysis of $B\to PP$ modes~\cite{Berthiaume:2023kmp}, which also includes the mixing-induced CP asymmetries, also found a very low \pvalue.
In contrast to this analysis, and as stated above, we do not impose the EWP-Tree relations in our analysis.
Imposing these relations in our setup can only worsen the fit's $\chi^2$.
We find, therefore, that the \SUF assumption is not favoured by the data even in the absence of such relations.

\begin{table}[ht]
    \centering
    \scalebox{0.95}{
        \begin{tabular}{lccc}
            \toprule
            \multirow{2}{*}{Channel} & \multicolumn{3}{c}{\textbf{Branching Ratios in units of $10^{-6}$}} \\[2pt]
             & \textbf{Experimental value} & \textbf{$\mathbf{\SUF}$} & \textbf{Fact.-$\cancel{\mathbf{\SUF}}$} \\
            \midrule
            $B^+\to \pi^+\pi^0$ & $5.31\pm 0.26$ &
            $5.35\pm 0.24$ & $5.33 \pm 0.24$ \\[0.4em]
            \multirow{2}{*}{$B^+\to K^+ \bar K^0$} & $1.32\pm 0.17$ \cite{LHCb:2013vip} &
            \multirow{2}{*}{$1.36\pm 0.14$} & \multirow{2}{*}{$1.41 \pm 0.14$} \\
            & $1.53 \pm 0.24^\dagger$ & & \\[0.4em]
            \multirow{2}{*}{${B}^0\to \pi^+ \pi^-$} & $5.37\pm 0.20$ &
            \multirow{2}{*}{$5.62 \pm 0.15$} & \multirow{2}{*}{$5.4 \pm 0.16$} \\
            & $5.24 \pm 0.40^ \dagger$ & & \\[0.4em]
            ${B}^0\to \pi^0 \pi^0$& $1.55\pm 0.17$&
            $1.49 \pm 0.14$ & $1.53 \pm 0.15$ \\[0.4em]
            ${B}^0\to K^+ K^-$& $0.079 \pm 0.015^\dagger$ &
            $0.091 _{-0.014}^{+0.013}$  & $0.075 _{-0.014}^{+0.015}$ \\[0.4em]
            ${B}^0\to \bar{K}^0 K^0$& $1.21 \pm 0.16$&
            $1.20_{-0.13}^{+0.15}$ & $1.22 \pm 0.16$ \\[0.4em]
            ${B}_s^0\to K^- \pi^+ $& $6.19 \pm 0.74^\dagger$&
            $6.18_{-0.21}^{+0.20}$ & $5.27 \pm 0.31$ \\[0.4em]
            ${B}_s^0\to \bar K^0 \pi^0 $& Not available&
            $1.14_{-0.15}^{+0.19}$& $1.37_{-0.16}^{+0.18}$ \\[0.4em]
            \midrule
            $B^+\to K^+ \pi^0$& $13.2\pm 0.4$&
            $13.00_{-0.33}^{+0.32}$ & $12.9\pm 0.34$ \\[0.4em]
            $B^+\to {K}^0 \pi^+$& $23.9\pm 0.6$&
            $24.09 \pm 0.53$ & $24.20 \pm 0.55$ \\[0.4em]
            ${B}^0\to K^+ \pi^-$& $20.0 \pm 0.4$&
            $19.60_{-0.35}^{+0.36}$ & $19.87 \pm 0.37$ \\[0.4em]
            ${B}^0\to {K}^0 \pi^0$& $10.1 \pm 0.4$&
            $10.36 \pm 0.31$ & $10.45 \pm 0.32$ \\[0.4em]
            ${B}_s^0\to \pi^+ \pi^-$& $0.766 \pm 0.096^\dagger$ &
            $0.706 \pm 0.085$& $0.756 \pm 0.090$ \\[0.4em]
            ${B}_s^0\to \pi^0\pi^0$& $2.8 \pm 2.8 \pm 0.5$ \cite{Belle:2023aau} &
            $0.353 \pm 0.043$ & $0.378_{-0.045}^{+0.046}$ \\[0.2em]
            \multirow{2}{*}{${B}_s^0\to K^+ K^-$} & $38 ^{+10}_{-9}\pm 7$ \cite{Belle:2010yix} &
            \multirow{2}{*}{$23.9_{-1.3}^{+1.1}$} & \multirow{2}{*}{$26.3 \pm 1.6$} \\
            & $26.4 \pm 2.0^\dagger$ & & \\[0.4em]
            ${B}_s^0\to K^0 \bar{K}^0$& $19.6^{+5.8}_{-5.1}\pm1.0 \pm 2.0$&
            $18.0_{-3.0}^{+2.8}$ & $17.5 \pm 3.1$ \\[0.4em]
            \bottomrule
        \end{tabular}
    }
    \caption{Experimental values and fit postdictions for $B\to PP$ branching ratios.
    Values without reference have been extracted from the PDG \cite{PDG2024}.
    Decays indicated with $\dagger$ have been measured using ratios relative to a control channel.
    While the ratio is used for the analysis, the branching ratio value is provided here for completeness.
    }
    \label{tab:table_topofit_BR}
\end{table}

\begin{table}[htp]
    \centering
    \scalebox{0.94}{
        \begin{tabular}{lccc}
        \toprule
        \multirow{2}{*}{Channel} & \multicolumn{3}{c}{\textbf{Ratios of Branching Ratios}} \\[2pt]
        & \textbf{Experimental value} & \textbf{$\mathbf{\SUF}$} & \textbf{Fact.-$\cancel{\mathbf{\SUF}}$} \\
        \midrule
        $\frac{f_s}{f_d}\frac{\mathcal{B}(B_s^0\to \pi^+\pi^-)}{\mathcal{B}(B^0\to K^+\pi^-)}$ & $\left(9.15 \pm 0.71 \pm 0.83 \right)\times 10^{-3}$ \cite{LHCb:2016inp} &
        $\left(8.6 \pm 1.0\right)\times 10^{-3}$ & $\left(9.1 \pm 1.1\right)\times 10^{-3}$ \\[0.4em]
        $\frac{f_s}{f_d}\frac{\mathcal{B}(B_s^0\to K^-\pi^+)}{\mathcal{B}(B^0\to K^+\pi^-)}$ & $0.074 \pm 0.006 \pm 0.006$ \cite{LHCb:2012ihl} &
        $0.0753 \pm 0.0025$ & $0.0634 \pm 0.0035$ \\[0.4em]
        $\frac{f_s}{f_d}\frac{\mathcal{B}(B_s^0\to K^+K^-)}{\mathcal{B}(B^0\to K^+\pi^-)}$ & $0.316 \pm 0.009 \pm 0.019$ \cite{LHCb:2012ihl} &
        $0.292_{-0.016}^{+0.013}$ & $0.317_{-0.020}^{+0.019}$ \\[0.4em]
        $\frac{\mathcal{B}(B^+\to \bar{K}^0K^+)}{\mathcal{B}(B^+\to K^0\pi^+)}$ & $0.064 \pm 0.009 \pm 0.004$ \cite{LHCb:2013vip} &
        $0.0567 \pm 0.0062$ & $0.0581 \pm 0.0057$ \\[0.4em]
        $\frac{\mathcal{B}(B^0\to \pi^+\pi^-)}{\mathcal{B}(B^0\to K^+\pi^-)}$ & $0.262 \pm 0.009 \pm 0.017$ \cite{LHCb:2012ihl} &
        $0.287 \pm 0.008$ & $0.2707 \pm 0.0088$ \\[0.4em]
        $\frac{\mathcal{B}(B^0\to K^+K^-)}{\mathcal{B}(B^0\to K^+\pi^-)}$ & $(3.98 \pm 0.65 \pm 0.42)\times 10^{-3}$\cite{LHCb:2016inp} &
        $\left(4.63_{-0.69}^{+0.67}\right)\times 10^{-3}$ & $\left( 3.78_{-0.68}^{+0.73}\right) \times 10^{-3}$ \\[0.4em]
        \bottomrule
        \end{tabular}
    }
    \caption{Experimental values and fit postdictions for ratios of $B\to PP$ branching ratios.
    The fragmentation fractions ratio is $f_s/f_d=0.239$~\cite{LHCb:2021qbv}.
    }
    \label{tab:table_topofit_ratBR}
\end{table}

\begin{table}[t]
    \centering
    \scalebox{0.95}{
        \begin{tabular}{lccc}
            \toprule 
            \multirow{2}{*}{Channel} & \multicolumn{3}{c}{\textbf{Direct CP asymmetries in units of $10^{-2}$}} \\[2pt]
            & \textbf{Experimental value} & \textbf{$\mathbf{\SUF}$} & \textbf{Fact.-$\cancel{\mathbf{\SUF}}$} \\
            \midrule
            $B^+\to \pi^+\pi^0$ & $1\pm 4$ &
            $1.2\pm 3.5$ & $1.3_{-3.6}^{+3.7}$ \\[0.4em]
            $B^+\to K^+ \bar K^0 $ & $9 \pm 10^*$ &
            $7.6 \pm 9.2$ & $9.7_{-10.2}^{+9.9}$ \\[0.4em]
            ${B}^0\to \pi^+ \pi^-$& {$-31.4\pm 3.0$}&
            $-36.2 \pm 2.2$ & $-34.2_{-1.9}^{+2.3}$ \\[0.4em]
            ${B}^0\to \pi^0 \pi^0$& $-30\pm 20$&
            $-32_{-13}^{+15}$ & $-30_{-16}^{+18}$ \\[0.4em]
            ${B}^0\to K^+ K^-$& Not available&
            $84_{-36}^{+12}$ & $-2_{-68}^{+73}$ \\[0.4em]
            $B^0\to \bar{K}^0 K^0$& $7 \pm 30^*$&
            $7_{-31}^{+30}$ & $7_{-30}^{+28}$ \\[0.4em]
            $B_s^0\to K^- \pi^+$& {$-22.4\pm 1.2 $}&
            $-24.50_{-0.90}^{+0.86}$ & $-21.9 \pm 1.0$ \\[0.4em]
            ${B}_s^0\to \bar K^0 \pi^0$&  Not available&
            $-53_{-12}^{+17}$ & $-38_{-17}^{+21}$ \\[0.4em]
            \midrule
            $B^+\to K^+ \pi^0$& {$-2.7\pm 1.2$}&
            $-2.6 \pm 1.2$ & $-2.6 \pm 1.2$ \\[0.4em]
            $B^+\to K^0 \pi^+$&$0.3 \pm 1.5$&
            $-0.43_{-0.55}^{+0.52}$ & $0.6 \pm 1.5$ \\[0.4em]
            ${B}^0\to K^+ \pi^-$& {$8.31\pm 0.31$}&
            $7.83 \pm 0.26$ & $8.41 \pm 0.29$ \\[0.4em]
            ${B}^0\to {K}^0 \pi^0$& $0\pm 8$&
            $5.9 \pm 2.3$ & $2.8 \pm 3.6$ \\[0.4em]
            ${B}_s^0\to \pi^+ \pi^-$& Not available&
            $-9.6_{-2.4}^{+4.0}$ & $0.1_{-4.7}^{+4.6}$ \\[0.4em]
            ${B}_s^0\to \pi^0\pi^0$& Not available&
            $-9.6_{-2.4}^{+4.0}$ & $0.1_{-4.7}^{+4.6}$ \\[0.4em]
            ${B}_s^0\to K^+ K^-$ & {$17.2\pm 3.1$} \cite{LHCb:2020byh} &
            $7.77_{-0.67}^{+0.71}$ & $11.0\pm 1.1$ \\[0.4em]
            ${B}_s^0\to K^0 \bar{K}^0$& Not available&
            $-0.4 \pm 1.9$ & $4_{-40}^{+37}$ \\[0.4em]
            \bottomrule
        \end{tabular}
    }
    \caption{Experimental values and fit postdictions for $B\to PP$ direct CP-asymmetries.
    Experimental results marked as ${}^*$ come from our own average (see \cref{sec:expinputs}), while the values without reference have been extracted from the PDG \cite{PDG2024}.
    }
    \label{tab:table_topofit_Adir}
\end{table}

\begin{table}[t]
    \centering
    \scalebox{0.95}{
        \begin{tabular}{lccc}
            \toprule 
            \multirow{2}{*}{Channel} & \multicolumn{3}{c}{\textbf{Mixing-induced CP asymmetries in units of $10^{-2}$}} \\[2pt]
            & \textbf{Experimental value} & \textbf{$\mathbf{\SUF}$} & \textbf{Fact.-$\cancel{\mathbf{\SUF}}$} \\
            \midrule
            ${B}^0\to \pi^+ \pi^-$& {$67 \pm 3 $}&
            $71.0_{-2.3}^{+2.2}$ & $66.8 \pm 2.1$ \\[0.4em]
            ${B}^0\to \pi^0 \pi^0$& Not Available &
            $-93.8_{-4.2}^{+6.6}$ & $-92.3_{-4.9}^{+9.3}$ \\[0.4em]
            ${B}^0\to K^+ K^-$& Not available&
            $-43_{-39}^{+49}$ & $-51_{-41}^{+97}$ \\[0.4em]
            $B^0\to \bar{K}^0 K^0$& $ 80 \pm 50 $&
            $80_{-33}^{+14}$ & $74_{-42}^{+20}$ \\[0.4em]
            ${B}_s^0\to \bar K_S^0 \pi^0$&  Not available&
            $45_{-16}^{+15}$ & $53_{-30}^{+25}$ \\[0.4em]
            \midrule
            ${B}^0\to {K}_S^0 \pi^0$& $-64 \pm 13 $&
            $-79.94_{-0.88}^{+0.92}$ & $-70.2_{-9.7}^{+19.6}$ \\[0.4em]
            ${B}_s^0\to \pi^+ \pi^-$& Not available&
            $4.9_{-5.7}^{+5.0}$ & $3.3_{-6.0}^{+3.8}$ \\[0.4em]
            ${B}_s^0\to \pi^0\pi^0$& Not available&
            $4.9_{-5.7}^{+5.0}$ & $3.3_{-6.0}^{+3.8}$ \\[0.4em]
            ${B}_s^0\to K^+ K^-$ & {$-13.9 \pm 3.2$}  \cite{LHCb:2020byh}&
            $-15.85_{-0.49}^{+0.42}$ & $-16.49_{-0.89}^{+0.69}$ \\[0.4em]
            ${B}_s^0\to K^0 \bar{K}^0$& Not available&
            $-4.8 \pm 2.3$ & $19 \pm 40$ \\[0.4em]
            \bottomrule
        \end{tabular}
    }
    \caption{Experimental values and fit postdictions for $B\to PP$ mixing-induced CP asymmetries.
    Values without reference have been extracted from the PDG \cite{PDG2024}.
    }
    \label{tab:table_topofit_Amix}
\end{table}

\subsubsection{Comment on the $\eta$ modes}
Our nominal analysis allows us to postdict the branching ratios and CP asymmetries for $B \to \eta P$ decays. We present our results in Appendix~\ref{app:eta}.
We observe that most modes agree with the measurements, except for the $B\to\eta K$ ones.

For completeness, we also performed the \SUF analysis, including the experimental data for the modes with $\eta$ mesons.
As discussed above, we assume $|\eta_8\rangle = |\eta\rangle$.
The fit quality is very poor, with a minimal $\chi^2 = 130$ for 23 degrees of freedom.
The tension at the best-fit point is mostly driven by $\mathcal{B}_\mathrm{exp}(B^{0,+}\to\eta K^{0,+})$, with pulls of approximately $7\sigma$ and $4\sigma$ for the neutral and charged modes respectively.

\section{\boldmath Factorizable \SUF breaking}\label{sec:factsuf}
To address the tensions found in the \SUF-limit, it is interesting to include \SUF-breaking corrections.
Completely relaxing the \SUF assumption increases the number of parameters beyond the number of observables.
The inclusion of linear \SUF-breaking effects by inserting the $m_s-m_d$ mass difference on the $s$-quark lines has been discussed in, e.g., Ref.~\cite{Gronau:1995hn}. 
This approach introduces a number of additional \SUF parameters whose magnitudes are, in principle, unknown.
We leave an update of this analysis based on the current experimental data for future work. 

An interesting and somewhat complementary option is to use the insights on the size of \SUF breaking from decay constants, form factors and the masses of the mesons.
These parameters enter into QCD factorization approaches, which factorize the amplitude in perturbatively calculable and non-perturbative hadronic parts.
Following such a factorization approach allows the inclusion of factorizable \SUF breaking corrections stemming from the differences in form factors and decay constants.
In the following, we perform a first analysis including factorizable \SUF-breaking effects for all $B\to PP$ decays combined.

To account for factorizable \SUF-breaking, we adopt the parametrisation used in QCD factorization (QCDF) \cite{Beneke:1999br,Beneke:2003zv}:
\begin{eqnarray}
   &&\hspace{-2.0cm}
   \mathcal{A}(\bar{B}_q\to M_1 M_2)=i\frac{G_{F}}{\sqrt{2}}\sum_{r=u,c}\,A^{B_q}_{M_1 M_2}\,\bigg\{
    B M_1 \left( \alpha_1\delta_{ru}\hat{U} +
    \alpha_4^r\hat{I}
     + \alpha_{4,EW}^r\hat{Q} \right) M_2\,
     \Lambda_r \nonumber\\
   &&\qquad\mbox{}+ B M_1 \Lambda_r\,
    \mbox{Tr}\left[\left( \alpha_2\delta_{ru} \hat{U} +  \alpha_{3,EW}^r\,\hat {Q} \right) M_2\right] \nonumber\\
   &&\qquad\mbox{}+ B \left( \beta_2\delta_{ru} \hat{U} + \beta_3^r\hat{I}
    + \beta_{3,EW}^r\,\hat{Q} \right)
    M_1 M_2\Lambda_r \nonumber\\
   &&\qquad\mbox{}+ B \Lambda_r\,
    \mbox{Tr}\left[\left( \beta_1 \delta_{ru} \hat{U} + \beta_4^r\hat{I}
    + \beta_{4,EW}^r\,\hat{Q} \right) M_1 M_2\right] \bigg\}\,,
    \label{eq:mastereq}
\end{eqnarray}
where we have omitted the singlet operators $\alpha_3, \beta_{S1}, \beta_{S2}, \beta_{S3, (\rm EW)}, \beta_{S4, (\rm EW)}$ since we do not consider $\eta^{(')}$ final states. In full generality, the coefficients $\alpha\equiv \alpha(M_1M_2)$ and $\beta \equiv \beta(M_1M_2)$ depend on the initial and final state mesons. Above,
\begin{align}
   \Lambda_r &= \left( \begin{array}{c}
    0 \\ \lambda_r^{(d)} \\ \lambda_r^{(s)}
   \end{array} \right), & \hat{U} &= \left( \begin{array}{ccc}
    1~ & 0~ & 0 \\
    0~ & 0~ & 0 \\
    0~ & 0~ & 0
   \end{array} \right) , & 
   \hat {Q} &= \frac{3}{2} Q =\frac{3}{2}\hat{U}-\frac{1}{2}\hat{I} = \left( \begin{array}{ccc}
    1 & 0 & 0 \\
    0 & -\frac12 & 0 \\
    0 & 0 & -\frac12
    \end{array} \right), 
\end{align}
and $\hat{I}$ is the identity matrix.
We note that $\beta_3$ can be absorbed into $\alpha_4$ due to their identical matrix structure.
Above, we already rewrote $\hat{Q}$ in terms of $\hat{U}$ and $\hat{I}$. This shows that the standard QCDF parametrisation in full generality, \ie without any \SUF assumptions, already has redundancies \cite{He:2018php}.
Specifically, we can redefine
\begin{eqnarray}\label{eq:redefs}
\tilde{\alpha}_1 \equiv \alpha_1 + \frac{3}{2}\alpha_{4,\text{EW}}^u,&~&
\tilde{\alpha}_2 \equiv \alpha_2 + \frac{3}{2}\alpha_{3,\text{EW}}^u,\nonumber\\
\tilde{\beta}_1 \equiv \beta_1 + \frac{3}{2}\beta_{4,\text{EW}}^u,&~&
\tilde{\beta}_2 \equiv \beta_2 + \frac{3}{2}\beta_{3,\text{EW}}^u, \nonumber\\
\tilde{\alpha}_{4}^r \equiv \alpha_4^r + \beta_3^r - \frac{1}{2}\left(\alpha_{4,\text{EW}}^r + \beta_{3,\text{EW}}^r\right), &~&  \tilde{\beta}_{4}^r \equiv \beta_4^r - \frac{1}{2} \beta_{4,\rm EW}^r,
\end{eqnarray}
This eliminates the electroweak parameters with CKM factor $\lambda_u$, while their $\lambda_c$ counterparts remain.
In total, this parametrisation has 12 independent complex parameters.

The factorizable \SUF-breaking effects are encoded in
\begin{equation}
A^{B_q}_{M_1 M_2}= M^2_B \, F^{B\to M_1}_0(m_{M_2}^2) \, f_{M_2},
\label{eq:AM1M2}
\end{equation}
where $M_B$ is the mass of the decaying $B$ meson, $F_0^{B\to M_1}(q^2)$ is the $B\to M_1$ form factor at $q^2$ and $f_{M_2}$ is the $M_2$ meson decay constant.
We explicitly show the flavour of the initial $B$ meson with a superscript. 

The $\beta_i$ parameters represent contributions from annihilation- and exchange-like topologies. As a result, the factorizable \SUF-breaking effects in these contributions are more accurately described by the decay constants of the mesons involved rather than form factors or $A^{B_q}_{M_1M_2}$.
In addition, for some pure annihilation modes, the parameter $A^{B_q}_{M_1M_2}$ cannot be defined because the form factors do not exist.
Consequently, we substitute all $\beta_i$ by $b_i$ using the following relation:
\begin{equation}
  A^{B_q}_{M_1M_2}  \beta_i^p \to B^{B_q}_{M_1M_2} \, b_i^p \hspace{1cm} \text{where} \hspace{1cm} B^{B_q}_{M_1M_2} = f_{B_q} f_{M_1} f_{M_2},
\end{equation}
where $f_{B_q}$ is the $B_q$ decay constant. 

In principle, additional \SUF breaking enters in the coefficients $\alpha(M_1M_2)$ and $\beta(M_1M_2)$. Accounting only for factorizable \SUF-breaking effects, boils down to the identifications $\alpha_i(M_1, M_2) \equiv \alpha_i$ and $b_i(M_1, M_2) \equiv b_i$.
Under this assumption, there is, for each decay, a one-to-one correspondence between the QCDF parametrisation and the topological parametrisation in \cref{sec:SUF_parametrisation}. This equivalence was pointed out already by Ref.~\cite{He:2018joe} and further discussed in Ref.~\cite{Huber:2021cgk}.
Unlike these references, our topological amplitudes are defined in the $(\lambda_{u},\lambda_c)$ basis, similar to the typical QCDF basis in Ref.~\cite{Beneke:2003zv}.
Therefore, we have 
\begin{align}
T&= A^{B_q}_{M_1M_2} \tilde\alpha_1, & C & = A^{B_q}_{M_1M_2} \tilde\alpha_2, & T_{P}&= A^{B_q}_{M_1M_2} \tilde\alpha_4^u,
\nonumber \\ 
A&= B^{B_q}_{M_1M_2} \tilde b_2, & E&= B^{B_q}_{M_1M_2} \tilde b_1, &
T_{PA}&= B^{B_q}_{M_1M_2} \tilde b_4^u, 
\label{eq:TopQCDFrelationTree}
\end{align}
and for the penguin parameters
\begin{align}
P_T&= \frac{3}{2}~A^{B_q}_{M_1M_2} \alpha_{4,EW}^c, & P_C & =\frac{3}{2} A^{B_q}_{M_1M_2} \alpha_{3,EW}^c, &
P&= A^{B_q}_{M_1M_2} \tilde\alpha_4^c, \nonumber\\  P_{TA}&= \frac{3}{2}B^{B_q}_{M_1M_2} b_{3,EW}^c, 
& P_{TE}&= \frac{3}{2}B^{B_q}_{M_1M_2} b_{4,EW}^c,&
P_A&= B^{B_q}_{M_1M_2} \tilde b_4^c.
\label{eq:TopQCDFrelationPeng}
\end{align}
where $\alpha_{4,EW}^c/P_T$ and $\alpha_{3,EW}^c/P_C$ respectively represent the color-suppressed and color-allowed electroweak penguin topologies. In the \SUF-symmetry limit, $A^{B_q}_{M_1 M_2}$ and $B^{B_q}_{M_1M_2}$ should be considered as a \textit{universal} constant as done in Ref.~\cite{Huber:2021cgk}. In this limit, we can directly convert our \SUF-limit analysis in the topological basis analysis of $\alpha,b$, setting also $\tilde{b}_1 = \tilde{b}_2$ and $b_{3,EW}^c=b_{4,EW}^c$ to account for the redundant parameters $E$ and $P_{TE}$ (see \cref{sec:SU3analysis:fitsetup}).  

Here, we proceed differently. The key point is that this parametrisation allows the inclusion of factorizable \SUF-breaking corrections.
In our setup, we take into account such corrections through initial and final state-dependent factors $A^{B_q}_{M_1M_2}$ and $B^{B_q}_{M_1M_2}$. 
Given the number of free parameters, we cannot consider the remaining non-factorizable \SUF dependence and thus maintain \SUF symmetry in the parameters $\alpha, b$.

To summarize, we parametrise the $\bar{B}\to M_1M_2$ decays as
\begin{align}
\label{eq:QCDFamp}
    \mathcal{A}_{\nSUF}(\bar{B}\to M_1M_2) &= A^{B_q}_{M_1M_2} \sum_{i=1}^3\left\{ \lambda_u^{(p)}\mathcal{T}^i_{\nSUF} + \lambda_c^{(p)}\mathcal{P}^i_{\nSUF}\right\} \nonumber \\
    & +  B^{B_q}_{M_1M_2} \sum_{i=4}^6\left\{ \lambda_u^{(p)}\mathcal{T}^i_{\nSUF} + \lambda_c^{(p)}\mathcal{P}^i_{\nSUF}\right\},
\end{align}
for a $b\to p$ transition with $p=d, s$.
In addition, 
\begin{align}\label{eq:ttop}
    \mathcal{T}_{\nSUF} &= \{\tilde\alpha_1, \tilde\alpha_2,  \tilde{\alpha}_4^u, \tilde{b}_2, \tilde b_1, \tilde{b}_4^u\}, \\
    \mathcal{P}_{\nSUF} &= \{\tfrac{3}{2}\alpha_{4,\rm EW}^c, \tfrac{3}{2}\alpha_{3,\rm EW}^c,  \tilde{\alpha}_4^c, \tfrac{3}{2}b_{3, \rm EW}^c, \tfrac{3}{2}b_{4,\rm EW}^c, \tilde{b}_4^c\}.
\end{align}
The $\mathcal{T}_{\nSUF}$ coefficients for each decay are given in \cref{tab:tableAmplBPPQCDF}.
The related penguin coefficients can be obtained by replacing $\mathcal{T}^i\to \mathcal{P}^i$. Topological coefficients can be obtained from \cref{tab:tableAmplBPPQCDF} as well by summing over the two possible configurations of the final state mesons.

We stress that, in the following, we do not make any assumptions on the size of $b$ and $\alpha$. 

\subsection[Factorizable \SUF-breaking]{\boldmath Factorizable \SUF-breaking}
Accounting for factorizable \SUF-breaking requires inputs on the form factors and decay constants.
We do not include isospin-breaking effects in these inputs and use the standard \EOS inputs.

In principle, the form factors should be evaluated at $q^2 = m_{M_2}^2$.
Using state-of-the-art form factor parametrisations~\cite{Boyd:1997kz,Gubernari:2023puw} to consider their kinematical dependence and reliably account for the form factor uncertainties would be numerically very expensive and would have little impact with respect to the current experimental uncertainties.
We assume, therefore, a simple parametrisation based on vector meson dominance
\begin{equation}
    F_0^{B\to M_1}(m_{M_2}^2) \simeq \frac{F_0^{B\to M_1}(0)}{1-m_{M_2}^2/m_{B_{p,0}}^2},
\end{equation}
where $m_{B_{p,0}}$ is the mass of the first scalar $\bar{b}p$ resonance.
For light mesons, this parametrisation differs from the more involved parametrisation in Ref.~\cite{Gubernari:2023puw} by less than 1\%.
For the current analysis, we do not vary $F_0^{B\to M_1}(0)$ for simplicity.
Due to the large experimental uncertainties, the effect of this simplification should be mild.

By definition, we have $B_{M_1 M_2} = B_{M_2 M_1}$, but for the $A$ factors, the order of the final state mesons matters. 
Taking the $B_d^0\to \pi^+\pi^-$ and $B_s^0 \to K^+K^-$ modes as an example, we find numerically
\begin{equation}
    A_{\pi \pi}^{B_d} = 1.11 \;{\rm GeV}^3 \ , \quad \quad A_{KK}^{B_s^0}  = 1.36\; {\rm GeV}^3 \ ,
\end{equation}
which results in $20\%$ \SUF breaking at the amplitude level for the $\alpha$ terms, which are expected to be dominant.
For the $\pi K$ modes, we find \SUF-breaking at the level of $7\%$ to $14\%$ from the $A^{B_q}_{M_1M_2}$ parameters. 

It is important to note that the numerical size of the $B_{M_1M_2}$ factor is very different from the $A_{M_1M_2}$ terms. Taking the pure annihilation modes as an example, we find numerically,
\begin{equation}
    B_{\pi \pi}^{B_s^0} = 3.9\times 10^{-3}\ {\rm GeV}^3 \ , \quad \quad  B_{KK}^{B_d} = 4.6\times 10^{-3}\ {\rm GeV}^3 \ ,
\end{equation}
which gives \SUF-breaking at the $15\%$ level. For comparison between the $\alpha$ and $b$ terms, we define: 
    \begin{equation}\label{eq:Cdef}
    C^{B_q}_{M_1M_2}  = \frac{B^{B_q}_{M_1M_2}}{A^{B_q}_{M_1M_2}} \ ,
\end{equation}
which is numerically around $\sim 3\cdot 10^{-3}$.

\begin{table}[t]
    \centering
    \renewcommand{\arraystretch}{1.25}
    \scalebox{0.85}{
        \begin{tabular}{lcccccc@{\hskip 15mm}lcccccc}
            \toprule 
            \textbf{$b\to d$ decay} & $\tilde \alpha_1$ & $\tilde \alpha_2$ & $\tilde \alpha_4^u$ & $\tilde b_2$ & $\tilde b_1$ & $\tilde b_4^u$ &     \textbf{$b\to s$ decay} & $\tilde \alpha_1$ & $\tilde \alpha_2$ & $\tilde \alpha_4^u$ & $\tilde b_2$ & $\tilde b_1$ & $\tilde b_4^u$ \\
            \midrule
            $B^+\to \pi^0\pi^+$ & $\frac{1}{\sqrt{2}}$ & $0$ & $\frac{1}{\sqrt{2}}$ & $\frac{1}{\sqrt{2}}$ & $0$ & $0$ & $B^+\to \pi^0 K^+$ & $\frac{1}{\sqrt{2}}$ & $0$ & $\frac{1}{\sqrt{2}}$ & $\frac{1}{\sqrt{2}}$ & $0$ & $0$ \\
            $B^+\to \pi^+\pi^0$ & $0$ & $\frac{1}{\sqrt{2}}$ & $-\frac{1}{\sqrt{2}}$ & $-\frac{1}{\sqrt{2}}$ & $0$ & $0$  &$B^+\to K^+ \pi^0$ & $0$ & $\frac{1}{\sqrt{2}}$ & $0$ & $0$ & $0$ & $0$ \\[3mm]
            $B^+\to \bar K^0 K^+$ & $0$ & $0$ & $0$ & $0$ & $0$ & $0$ &  $B^+\to K^0 \pi^+ $ & $0$ & $0$ & $0$ & $0$ & $0$ & $0$ \\
            $B^+\to K^+ \bar K^0 $ & $0$ & $0$ & $1$ & $1$ & $0$ & $0$  & $B^+\to \pi^+ K^0$ & $0$ & $0$ & $1$ & $1$ & $0$ & $0$   \\[3mm]
            $B^0\to \pi^+ \pi^-$ & $0$ & $0$ & $0$ & $0$ & $1$ & $1$ &   $B_s^0\to K^+ K^-$ & $0$ & $0$ & $0$ & $0$ & $1$ & $1$  \\
            $B^0\to \pi^- \pi^+$ & $1$ & $0$ & $1$ & $0$ & $0$ & $1$ &    $B_s^0\to K^- K^+ $ & $1$ & $0$ & $1$ & $0$ & $0$ & $1$    \\[3mm]
            $B^0\to \pi^0 \pi^0$ & $0$ & $-1$ & $1$ & $0$ & $1$ & $2$  & $B_s^0\to \pi^0\pi^0$ & $0$ & $0$ & $0$ & $0$ & $1$ & $2$   \\[3mm]
            $B^0\to K^+ K^-$ & $0$ & $0$ & $0$ & $0$ & $1$ & $1$ & $B_s^0\to \pi^+ \pi^-$ & $0$ & $0$ & $0$ & $0$ & $1$ & $1$\\
            $B^0\to K^- K^+$ & $0$ & $0$ & $0$ & $0$ & $0$ & $1$  &$B_s^0\to \pi^- \pi^+$ & $0$ & $0$ & $0$ & $0$ & $0$ & $1$    \\[3mm]
            $B^0\to K^0 \bar{K}^0$ & $0$ & $0$ & $1$ & $0$ & $0$ & $1$ & $B_s^0\to \bar{K}^0 K^0$ & $0$ & $0$ & $1$ & $0$ & $0$ & $1$  \\
            $B^0\to \bar{K}^0 K^0 $ & $0$ & $0$ & $0$ & $0$ & $0$ & $1$ & $B_s^0\to K^0 \bar{K}^0$ & $0$ & $0$ & $0$ & $0$ & $0$ & $1$    \\[3mm]
            $B_s^0\to \pi^+ K^-$ & $0$ & $0$ & $0$ & $0$ & $0$ & $0$ & $B^0\to K^+ \pi^-$ & $0$ & $0$ & $0$ & $0$ & $0$ & $0$  \\
            $B_s^0\to K^- \pi^+ $ & $1$ & $0$ & $1$ & $0$ & $0$ & $0$ &$B^0\to \pi^- K^+$ & $1$ & $0$ & $1$ & $0$ & $0$ & $0$    \\[3mm]
            $B_s^0\to \pi^0 \bar K^0$ & $0$ & $0$ & $0$ & $0$ & $0$ & $0$ &   $B^0\to \pi^0  K^0$ & $0$ & $0$ & $-\frac{1}{\sqrt{2}}$ & $0$ & $0$ & $0$ \\
            $B_s^0\to \bar K^0 \pi^0$ & $0$ & $\frac{1}{\sqrt{2}}$ & $-\frac{1}{\sqrt{2}}$ & $0$ & $0$ & $0$ &    $B^0\to K^0 \pi^0 $ & $0$ & $\frac{1}{\sqrt{2}}$ & $0$ & $0$ & $0$ & $0$ \\
            \bottomrule
        \end{tabular}
    }
    \caption{Tree parameters $\mathcal{T}_i$ contributing to the $B\to PP$ decay in \eqref{eq:QCDFamp}.
    The penguin components are obtained by $\mathcal{T}_i\to \mathcal{P}_i$ using \eqref{eq:ttop}. Coefficients are in agreement with \cite{Beneke:2003zv}.} 
    \label{tab:tableAmplBPPQCDF}
\end{table}

\section{\boldmath Analysis including Factorizable \SUF breaking}
We proceed by setting up the analysis with factorizable \SUF-breaking corrections.
These corrections break the one-to-one correspondence with the \SUF decomposition, and we can no longer remove the parameters associated with $E$ and $P_{TE}$ as in the topological analysis.
Removing the arbitrary overall phase leaves 23 real parameters. 

We perform a purely data-driven analysis without accounting for form factors and decay constants uncertainties.
At the best-fit point, we obtain an excellent fit with a \pvalue of $32.3\%$.
A goodness-of-fit summary is provided in~\cref{tab:gof}, where we also compare it to the \SUF-symmetric fit.
We conclude that the $20$--$30\%$ \SUF-breaking from the factorizable form factors and decay constants allows for a perfect description of the current $B_{(s)}\to \pi K, KK$ and $\pi\pi$ data.
Our conclusion seems to be in contrast to \cite{Berthiaume:2023kmp}, where $\Delta S=0$ and $\Delta S=1$ decays were considered separately, and it was claimed that effects of $1000\%$ \SUF breaking (for certain topologies) are required to understand the data.

\begin{table}[t]
    \centering
    \begin{tabular}{lcccc}
        \toprule
         & $\chi^2$ & constraints & parameters & \pvalue [\%]\\
        \midrule
        \SUF                     & 32.3 & 34 & 19 & 0.58 \\
        Fact.-$\cancel{{\SUF}}$  & 12.6 & 34 & 23 & 32.3 \\
        \bottomrule
    \end{tabular}
    \caption{Goodness-of-fit summary of our analyses. }
    \label{tab:gof}
\end{table}

We provide postdictions for all our observables in \cref{tab:table_topofit_BR,tab:table_topofit_Amix,tab:table_topofit_Adir,tab:table_topofit_ratBR,tab:table_topofit_ADG}, and in \cref{fig:fit_noeta}.
We note that unmeasured observables have larger uncertainties within this parametrisation than under the \SUF symmetry assumption.
This is explained by the poor fit quality of the very constrained \SUF analysis, which biased the fit results.

\subsection{Constraints on parameters}
The parameter's posterior distributions show strong correlations associated with several poorly constrained directions.
Due to the non-linearity of the constraints, the posterior distributions are again non-Gaussian. 
The appearing poorly constrained directions can be understood from Table~\ref{tab:tableAmplBPPQCDF}:
\begin{itemize}
\item First and most importantly, $\tilde\alpha_1$ only appears in the combination $\tilde\alpha_1 + \tilde\alpha_4^u$.
We fix, therefore, the global phase by setting $\tilde\alpha_1 + \tilde\alpha_4^u >0$.
\item Second, $\tilde{\alpha}_2$ appears in most decays in the combination $\tilde\alpha_2 - \tilde\alpha_4^u$.
Only the factorizable \SUF-breaking corrections in $B^+\to \pi^0 K^+$ and $B^0\to \pi^0 K^0$ disentangle the two contributions.
\item In addition, from the annihilation operators, only the combination $\tilde b_1 + 2\tilde b_4^u$ appears.
\item Finally, we also note that no decay constrains only $\alpha_4^u$; this parameter always comes with weak-annihilation parameters $\tilde{b}_2$ or $\tilde{b}_4^u$, which we discuss in Sec.~\ref{sec:weakpluspenguin}.
\end{itemize}
These observations also hold for the penguin-counterparts of these parameters. 
The combinations discussed above are very well constrained by the data.
We find
\begin{equation}\label{eq:a1a4}
    \tilde\alpha_1 + \tilde\alpha_4^u = 0.584 \pm 0.024 \ ,
\end{equation}
and for their penguin counterparts
\begin{equation}
   \frac{3}{2} \alpha_{4,\rm EW}^c + \tilde\alpha_4^c = -\big(0.102 \pm 0.001\big) + \big(0.044 \pm 0.002\big) \,i,
\end{equation}
which are strongly constrained by the $U$-spin partners $B_s^0\to \pi^+K^-$ and $B_d^0\to K^+\pi^-$. 

In addition,
\begin{equation}
    \tilde\alpha_2 - \tilde\alpha_4^u = \big( 0.414^{+0.053}_{-0.065} \big) - \big(0.34 \pm 0.11\big) \,i,
    \quad |\tilde\alpha_2 - \tilde\alpha_4^u| = 0.539_{-0.038}^{+0.041} \ ,
\end{equation}
and for the penguin counterparts
\begin{equation}
    \frac{3}{2}\alpha_{3,EW}^c - \tilde\alpha_4^c = \big(0.141_{-0.024}^{+0.031}\big) - \big(0.059 \pm 0.010\big) \,i,
\end{equation}
which are mainly driven by the $B_d^0\to K^0 \pi^0$ decay. 

\begin{figure}[t]
    \centering
    \includegraphics[width=0.99\textwidth]{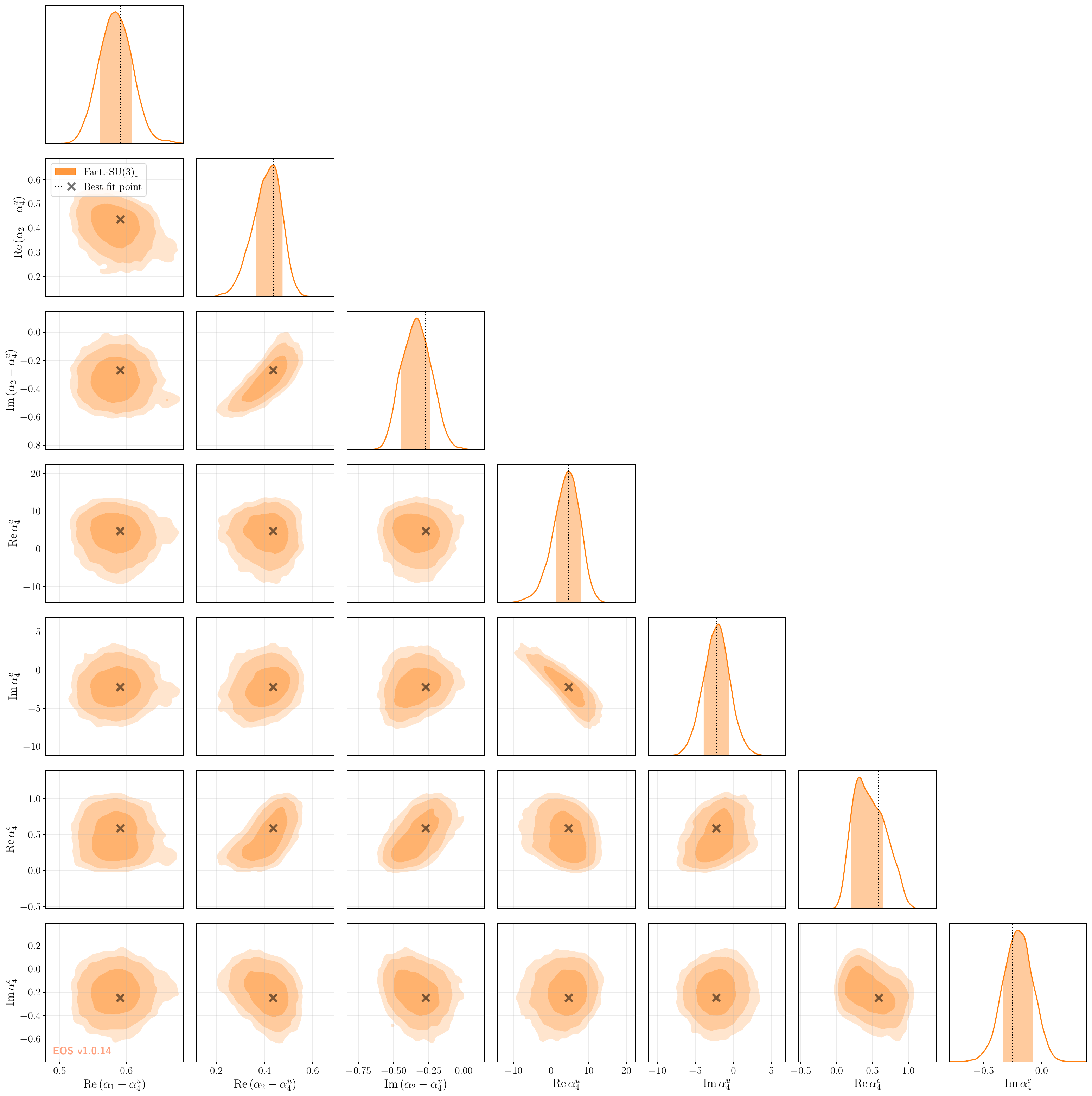}
    \caption{
        Corner plot of one-dimensional and two-dimensional marginalised posterior densities for the several $\alpha$ contributions.
        The ``$\times$'' symbol and the dashed black lines show the analysis best-fit point.
        The orange areas are the 1, 2 and $3\sigma$ contours of the posterior distribution obtained from a kernel density estimation.
    }
    \label{fig:cornerplot}
\end{figure}

\Cref{fig:cornerplot} shows the posterior distributions of the dominant $\alpha$ contributions and their best-fit value.
The complete corner plot of all the posterior distributions is provided in the supplementary material~\cite{EOS-DATA-2025-01}.

Due to the poorly constrained directions in the fit, the distributions of the individual parameters cover wide ranges.
Specifically, comparing the marginalised posterior of $\tilde\alpha_1$ with the result of QCD Factorization is uninformative.
Furthermore, the calculation of some of these parameters in QCDF is challenging.
For example, due to the parameter redefinitions described in eq.~\eqref{eq:redefs}, $\tilde{\alpha}_4$ contains also weak-annihilation contributions $\beta_{3, (\rm EW)}$, which can only be modelled.
At the same time, we note that the factorizable \SUF-breaking correction breaks, in principle, the redundancy in the parametrisation such that $\beta_3$ should no longer be re-absorbed in $\alpha_4$.
However, at the moment, the data have only limited constraining power on $\tilde{\alpha}_4^u$ and $\tilde{\alpha}_4^c$.
As such, we conclude that distinguishing $\beta_3$ from $\alpha_4$ would require more precise data on the penguin-dominated modes.
We leave a detailed investigation of this point for future work. 

\begin{figure}[t]
    \centering
    \includegraphics[width=0.5\textwidth]{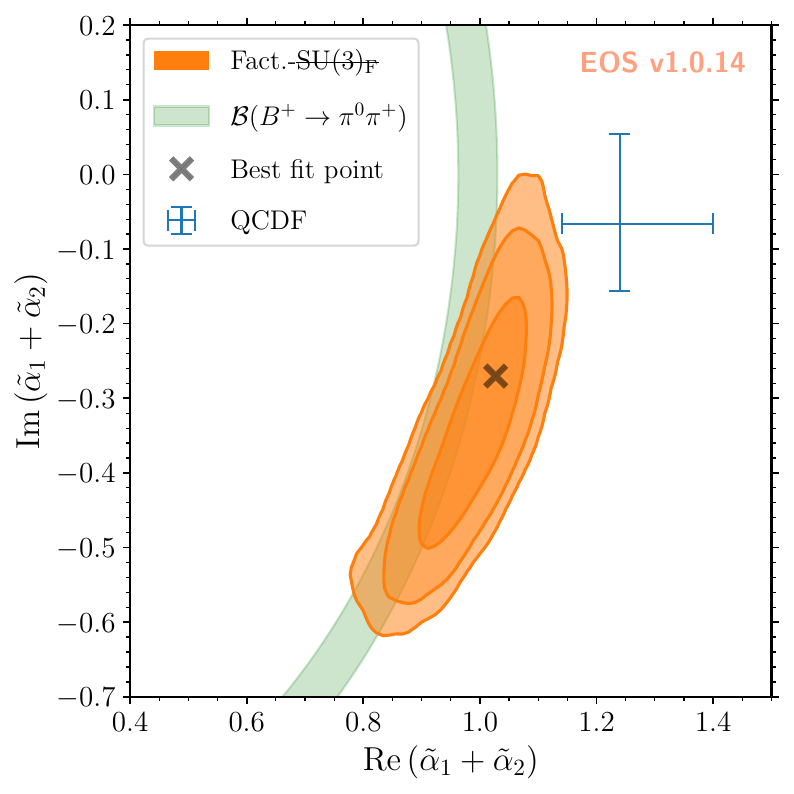}
    \caption{
        68\%, 95\% and 99\% cumulative contours of the $\tilde\alpha_1 + \tilde\alpha_2$ posterior distribution, assuming $\tilde\alpha_1 + \alpha_4^u > 0$.
        The ``$\times$'' symbol shows the analysis best-fit point.
        We overlay the QCDF result for $\alpha_1 + \alpha_2$ obtained in Ref.~\cite{Beneke:2009ek}.
    }
    \label{fig:alpha1andalpha2}
\end{figure}

Given the discussion above, $\tilde\alpha_1+\tilde\alpha_2$ is the cleanest combination to compare our results to QCDF.
We show the real and imaginary parts of this observable in Fig.~\ref{fig:alpha1andalpha2} and remind that we have constrained \eqref{eq:a1a4} to be real and positive.
In principle, this means that we only constrain the phase difference of our parameters with respect to phase in $\tilde\alpha_1+\tilde\alpha_4^u$.
Neglecting electroweak parameters $\alpha_{4,\rm EW}^c$ and $\alpha_{3,\rm EW}^c$, the $B^+\to \pi^0\pi^+$ branching ratio directly constrains\footnote{%
This value is lower than that obtained in Ref.~\cite{Beneke:2009ek}, due to a $1\sigma$ downward shift in the experimental branching ratio and a larger value for $|V_{ub}F^{B\to \pi}|$. 
}
\begin{equation}\label{eq:exp}
    |\tilde\alpha_1 + \tilde\alpha_2|_{B^+\to \pi^0\pi^+} = 1.03\pm 0.03 \pm 0.06\ , 
\end{equation}
where the first is the experimental uncertainty and the second a conservative theoretical uncertainty stemming from the input for $|V_{ub}F^{B\to \pi}|$. 
QCDF predicts at NNLO $|\alpha_1+\alpha_2|_\mathrm{QCDF} = 1.24^{+0.16}_{-0.10}$ \cite{Beneke:2009ek}.
Neglecting the electroweak parameters in our redefined $\tilde{\alpha}_1$ and $\tilde\alpha_{2}$, which are expected to be suppressed, we find a $2\sigma$ tension with our results.
These constraints are also shown in Fig.~\ref{fig:alpha1andalpha2}, where we neglect the phase in $\tilde\alpha_1+\tilde\alpha_4^u$ for comparison.
Within QCDF, this phase is found to be $\sim 3^\degree$ \cite{Beneke:2009ek, Bell:2015koa}.

For the other $\alpha$ parameters, we find
\begin{equation}
    \left|\frac{\alpha_{4,EW}^c}{\tilde\alpha_1}\right| \sim
    \left|\frac{\tilde\alpha_4^c}{\tilde\alpha_4^u}\right| \sim
    \left|\frac{\alpha_{3,EW}^c}{\tilde\alpha_2}\right| \in [10^{-4}, 10^{-2}] \,,
\end{equation}
or more precisely, all the logarithms of these ratios $r_i$ are Gaussian-distributed with $\log{r_i}\sim -3\pm 1$.
Surprisingly, the $c$-penguin coefficient $\tilde\alpha_4^c$ is much smaller than its $u$ counterpart.
We also find that the electroweak corrections can be neglected with respect to $\tilde\alpha_1$ and $\tilde\alpha_2$. 
However, we find 
\begin{equation}\label{eq:eweak}
    \left|\frac{\alpha^c_{4,{\rm EW}}}{\tilde\alpha_4^c}\right| \sim 1 \ ,
\end{equation}
which is much larger than what is typically expected due to the suppression of the electroweak operators (see e.g.~\cite{Neubert:1998jq, Beneke:2003zv}). Our finding that the electroweak contributions cannot be neglected with respect to the penguin parameters is important to remember in the phenomenological discussion below. 

For the weak-annihilation parameters $b_i$, we can strongly constrain the combination
\begin{equation}\label{eq:b1b4tree}
    \tilde{b}_1 + 2\tilde b_4^u = -\big(3.7^{+9.6}_{-8.2}\big) + \big(13.7^{+6.9}_{-11.9}\big) \,i,
\end{equation}
and its penguin counterparts 
\begin{equation}\label{eq:b1b4pen}
    \frac{3}{2}b_{4,EW}^c + 2\tilde b_4^c = -\big(1.5_{-1.5}^{+1.4}\big) + \big(11.0_{-5.2}^{+4.6}\big) \,i.
\end{equation}
These constraints are dominated by the data on the $B^0_d \to K^+K^-$ decay and its $U$-spin partner $B_s^0\to \pi^+ \pi^-$.
Comparing the size of these parameters to the $\alpha$ parameters requires taking into account their respective prefactors through $C_{M_1M_2}^{B_q}$ defined in \eqref{eq:Cdef} which is numerically of order $\mathcal{O}(3\cdot 10^{-3})$. 

In addition, we find
\begin{equation}
    \left|\frac{\tilde b_2}{\tilde b_1}\right| = 0.99 \pm 0.05 .
\end{equation}
Finally, as with the $\alpha$ parameters, we obtain 
\begin{equation}
    \left|\frac{b_{4,EW}^c}{\tilde b_1}\right| \sim
    \left|\frac{b_{3,EW}^c}{\tilde b_2}\right| \sim
    \left|\frac{\tilde b_4^c}{\tilde b_4^u}\right| \in [10^{-4}, 10^{-2}] \,,
\end{equation}
indicating that the electroweak parameters are small and that $\tilde b_4^c$ is much smaller than $\tilde b_4^u$. 

We note again that, due to poorly constrained directions, all the individual parameters of the analysis can vary in broad ranges and are strongly correlated.

\subsection{Phenomenological results}
Our analysis would benefit from more - or more precise - experimental data to test our assumptions further.
However, we can already predict modes that have not yet been observed experimentally.
Additionally, for several modes, we can make more precise postdictions than the current experimental data.
Below, we highlight some of the key phenomenological outcomes of our analysis to guide the experimental program.

\subsubsection[$B_d^0\to \pi^+\pi^-$ and $B_s^0 \to K^+K^-$ modes]{\boldmath $B_d^0\to \pi^+\pi^-$ and $B_s^0 \to K^+K^-$ modes}
In \cref{sec:su3_SR}, we found that the CP asymmetries in these modes were in tension with the strict \SUF limit. To show how factorizable \SUF-breaking enters, we express the ratio in \eqref{eq:rat} in our new parametrisation:
\begin{equation}\label{eq:rppipi}
    r_d e^{i\theta_d} = \frac{\tilde{\alpha}_4^c + \frac{3}{2}\alpha_{4,\rm EW}^c + C^{B_d}_{\pi \pi}(\frac{3}{2}{b}_{3,\rm EW}^c+ 2\tilde{b}_4^c)}{\tilde{\alpha}_1+\tilde{\alpha}_4^u + C^{B_d}_{\pi \pi}(\tilde{b}_1+ 2\tilde{b}_4^u)} \ , \quad     r_s e^{i\theta_s} = \frac{\tilde{\alpha}_4^c + \frac{3}{2}\alpha_{4,\rm EW}^c  + C^{B_s}_{KK}(\frac{3}{2}{b}_{3,\rm EW}^c+ 2\tilde{b}_4^c) }{\tilde{\alpha}_1+\tilde{\alpha}_4^u + C^{B_s}_{KK}(\tilde{b}_1+ 2\tilde{b}_4^u)} \ ,
\end{equation}
where $C$ is defined in \eqref{eq:Cdef}.
We thus note that neglecting the annihilation topologies gives $r_s=r_d$ and we recover the \SUF-limit relation.

We remind that in \cref{sec:su3_SR}, $r_s$ and $r_d$ were extracted from the CP asymmetries alone without any assumptions, where we found $|r_s/r_d| \sim 0.9$.
A quick estimate shows that $b \sim \mathcal{O}(10)$ effects can easily account for that amount of \SUF breaking, indicating that including factorizable \SUF-breaking can remove the tension observed in the \SUF limit. 

From our full analysis, we still observe small tensions in the $B_s^0 \to K^+K^-$ CP asymmetries.
Especially the direct CP asymmetry
\begin{equation}
    \mathcal{A}^{\rm dir}_{\rm CP}(B_s^0\to K^+ K^-) = (11.0 \pm 1.1)\% \ ,
\end{equation}
is in $1.7\sigma$ tension with the experimental result by the LHCb collaboration \cite{LHCb:2020byh} and has a factor 3 smaller uncertainty.
This measurement was the first observation of CP violation in $B_s^0$ decays.
It would be interesting to compare our analysis with future, more precise measurements of this quantity.
The mixing-induced CP asymmetry is consistent with the current experimental results, and our postdiction has an impressive uncertainty of about 0.5\%.
Additionally, our postdictions for the $U$-spin partner mode, $B_d^0 \to \pi^+ \pi^-$, are in perfect agreement with experimental data.
Finally, for the branching ratio, we find good agreement with the measured ratio between the $B_s^0$ and $B_d^0$ modes.

We conclude that even though these decays are amongst the most precisely measured $B\to PP$ modes, updates of these two, tree-dominated modes would still be important to further these \SUF-test of the non-leptonic decays.

\subsubsection[$B_{(s)}^0\to \pi K$ modes]{\boldmath $B_{(s)}^0\to \pi K$ modes}
The CP asymmetries and the ratio of the branching ratios of the $U$-spin partner modes $B_s^0\to \pi^+K^-$ and $B_d^0\to K^+\pi^-$ can be used to directly constrain the ratio of the $A^{ct}$ over $A^{ut}$ or penguin-like topologies over tree-like topologies, defined in \eqref{eq:rat}. In our parametrisation, we then have 
\begin{equation}
    \left. r_s e^{i\theta_s} \right|_{B_d^0\to K^+ \pi^-} =
    \left. r_d e^{i\theta_d} \right|_{B_s\to \pi^+ K^-}  =  \frac{\tilde\alpha_4^c+ \frac{3}{2}\alpha_{4,\rm EW}^c}{\tilde\alpha_1 + \tilde\alpha_4^u} \ ,
\end{equation}
where we indicate which decay we consider to distinguish from the $(r_p, \theta_p)$ parameters in \eqref{eq:rppipi}. We note that for these decays, the factorizable contributions drop out in the ratio and the two parameters are equal. For the branching ratio, the factorizable contributions add an over-all factor. 
\begin{figure}[t]
    \centering
    \includegraphics[width=0.8\textwidth]{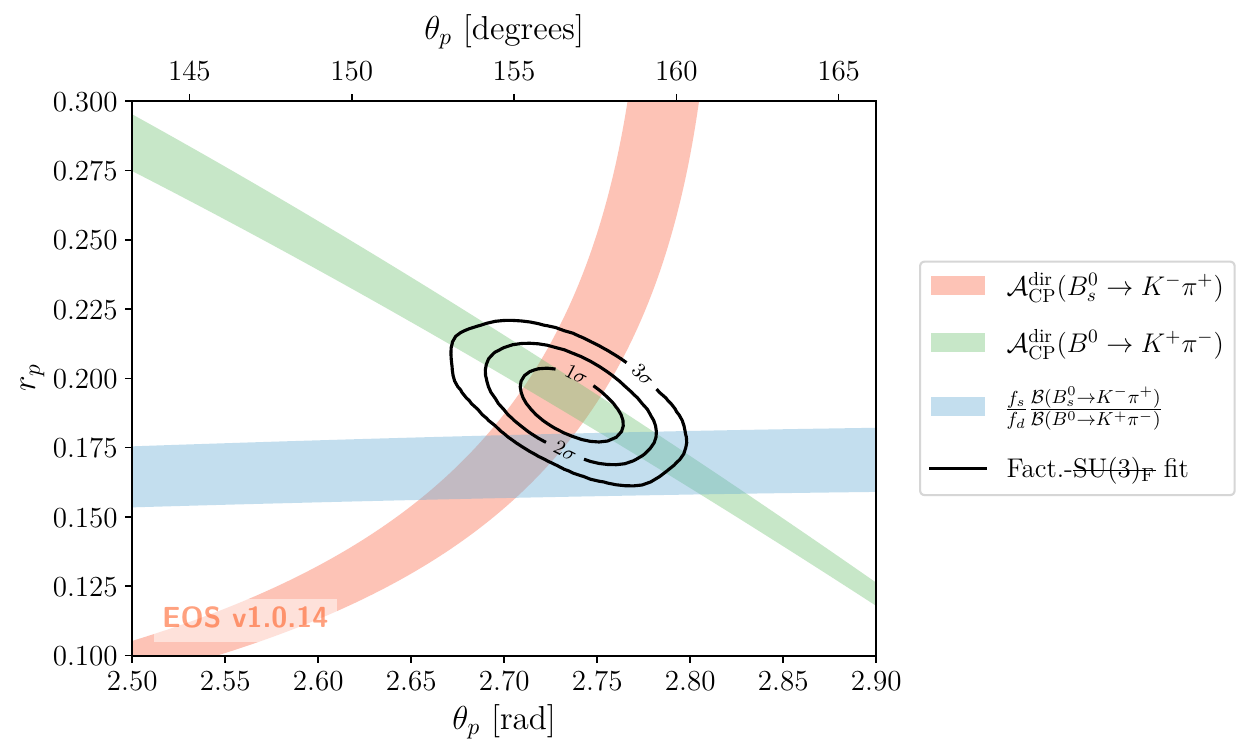}
    \caption{
        $68\%$ probability intervals of the (partially correlated) experimental constraints on the $B_d^0\to K^+\pi^-$ and $B_s^0\to \pi^+K^-$
        CP observables and ratio of branching ratio in the $(r, \theta)$ plane defined in \eqref{eq:rat}.
        The $1, 2$ and $3 \sigma$ postdictions of our fit are overlaid.
    }
    \label{fig:pik}
\end{figure}
In Fig.~\ref{fig:pik}, we show the constraints on $(r_p,\theta_p)$ from the current data, which has an impressive $5\%$ uncertainty for the CP asymmetries. These decays were also studied in detail in e.g.~\cite{Fleischer:2017vrb, Fleischer:2018bld}. We observe a small tension between the three bands.
The result of our factorizable \SUF-breaking global fit is overlaid, indicating that we can accommodate the data.
This also indicates that non-factorizable contributions, which would introduce a difference between $r_s$ and $r_d$, are small. 

Our analysis predicts small shifts in the CP asymmetries to compensate for the small mismatch between these three bands. The largest predicted shift is in the ratio of the branching ratios, which we predict to go down by about $1\sigma$. This ratio is only available from the LHCb collaboration, and requires input for the production fraction $f_s/f_d$. An updated analysis of this quantity would therefore be of interest.

Similarly, CP asymmetries of the $B_s^0\to \pi^0 \bar{K}^0$ and $B_d^0\to K^0 \pi^0$ $U$-spin pair constrain
\begin{equation}
   \frac{\tilde\alpha_4^c + \frac{3}{2}\alpha_{3,\rm EW}^c}{\tilde\alpha_2 - \tilde\alpha_4^u} \ .
\end{equation}
However, only a first measurement of the direct CP asymmetries of the $B_d^0$ mode is available, while no measurements are available for the $B_s^0$ mode.
The power of our combined analysis is that we can predict observables in these modes with an improved accuracy.
We find
\begin{equation}
    \mathcal{B}(B_s^0\to \pi^0\bar{K}^0) = \left( 1.37^{+0.18}_{-0.16} \right) \times 10^{-6}
\end{equation}
and
\begin{equation}
\mathcal{A}^{\rm dir}_{\rm CP}(B_d^0\to K^0 \pi^0)= (2.8 \pm 3.6)\% \ , \qquad \mathcal{A}^{\rm dir}_{\rm CP}(B_s^0\to \pi^0 \bar{K}^0) = \left(-38^{+21}_{-17} \right)\% \ .
\end{equation}
Experimental confirmation of these values would allow to further understand these decays. 

In general, it is challenging to predict the observables for the $B\to \pi K$ decays due to annihilation effects.
It is therefore interesting to consider the sum rule
\cite{Gronau:2005kz, Gronau:2006xu}:
\begin{align}\label{eq:SR}
\Delta_{\rm SR} &= 	-    \mathcal{A}^{\rm{dir}}_{\rm{CP}}(B^+\to \pi^0K^+) \frac{2 {\rm  \mathcal{B}}(B^+\to\pi^0 K^+)}{{\mathcal{B}}(B_d^0\to \pi^- K^+)} \frac{\tau_{B^0}}{\tau_{B^+}} -   \mathcal{A}^{\rm{dir}}_{\rm{CP}}(B_d^0\to \pi^0 K^0) \frac{2  {\mathcal{B}}(B_d^0 \to \pi^0 K^0)}{{\mathcal{B}}(B_d^0 \to \pi^- K^+)} \nonumber \\
& + \mathcal{A}^{\rm{dir}}_{\rm{CP}}(B_d^0 
\to \pi^- K^+)  +   \mathcal{A}^{\rm{dir}}_{\rm{CP}}(B^+ \to \pi^+ K^0) \frac{{ \mathcal{B}}(B^+ \to \pi^+ K^0)}{{\mathcal{B}}(B_d^0 \to \pi^- K^+)} \frac{\tau_{B^0}}{\tau_{B^+}} 
\ ,						
\end{align}
which is expected to be small as all linear hadronic effects in terms of $\alpha$ and $b$ cancel. 
In QCDF, we have $\Delta_{\rm SR}^\mathrm{QCDF} \sim 1\%$~\cite{Bell:2015koa, Beneke:2020vnb}, which includes an estimate for the weak-annihilation effects. 

Using the experimental data in~\cref{tab:table_topofit_BR,tab:table_topofit_Adir}, we find
\begin{equation}
    \Delta_{\rm SR}^\mathrm{exp.} = 0.12 \pm 0.08 \ ,
\end{equation}
which does not take the (unknown) correlations of the PDG average results into account and still has a sizeable uncertainty. For completeness, we also quote that the Belle II collaboration recently measured all the inputs of this sum rule, including experimental correlations.
They find $\Delta_{\rm SR}^\mathrm{Belle II} = 0.03 \pm 0.13\pm 0.04$~\cite{Belle-II:2023ksq}, which is consistent with theoretical expectations. 
We can also predict this sum rule from our global analysis. We find 
\begin{equation}
    \Delta_{\rm SR}^{\mathrm{Fact.} \cancel{{\rm SU(3)}_{\rm F}}} = 0.097 \pm 0.050 \ ,
\end{equation}
which has a sizeable uncertainty but agrees with the general theoretical expectations.  

We eagerly await further updates on these modes, specifically from the Belle II collaboration on the neutral $\pi$ modes, to further confirm our factorizable \SUF assumption. 

In the context of the $B\to \pi K$ decays, often the difference in the CP asymmetries between the $B^+\to \pi^0 K^+$ and $B^0\to \pi^- K^+$ is discussed.
As we do not observe any large tensions in the $B\to \pi K$ modes, we find
\begin{equation}
    \delta(\pi K) =\mathcal{A}_{\rm CP}^{\rm dir}(B^+\to \pi^0 K^+) - \mathcal{A}_{\rm CP}^{\rm dir}(B^0\to \pi^- K^+) = \left(-11.0^{+1.2}_{-1.3}\right) \%\ ,
\end{equation}
which is in perfect agreement with the experimental value $\delta(\pi K)|_{\rm exp}= (-11.0 \pm 1.2)\%$.
This is at odds with the QCDF calculations, which predict $\delta(\pi K)\sim (0-5)\%$ \cite{Bell:2015koa}.
This difference may be understood from our finding that the electroweak contributions are of similar size as the QCD penguin coefficients as shown in \eqref{eq:eweak}.
We will defer a more detailed discussion of these unexpectedly large electroweak effects to a forthcoming publication. 

In conclusion, we note that in our analysis, we do not observe large tensions in the $\pi K$ modes, and factorizable \SUF-breaking can perfectly accommodate the data.

\subsubsection[$B^0_{(s)}\to \bar{K}^0 K^0$ and $(B^+\to K^0 K^+, B^+\to K^0 \pi^+)$ modes]{\boldmath $B^0_{(s)}\to \bar{K}^0 K^0$ and $(B^+\to K^0 K^+, B^+\to K^0 \pi^+)$ modes}\label{sec:weakpluspenguin}
The $B^0_{(s)}\to \bar{K}^0 K^0$ decays and $B^+\to K^0 K^+,B^+\to K^0\pi^+$ decays modes probe the combination of $\tilde\alpha_4^u$ with the weak-annihilation parameters $\tilde b_4^u$ and $\tilde b_2$, respectively.
For all these four modes, we exactly reproduce the available data with limited constraining power.
This can be understood because these modes have a unique sensitivity to $\tilde\alpha_4^u$, which otherwise only occurs in combination with $\tilde\alpha_1$ or $\tilde\alpha_2$.
As such, they are key to further distinguishing the $\tilde\alpha_4^u$ parameter from the tree parameters $\tilde\alpha_1$ and $\tilde\alpha_2$ and distinguishing the strength of weak-annihilation versus penguin parameter.
In general, the effect of the $\tilde b_4$ and $\tilde b_2$ parameters is expected to be suppressed with respect to the penguin parameter due to the suppression by the prefactors given by $C_{M_1M_2}^{B_q}$, which is of order $10^{-3}$.
However, we find
\begin{equation}
  \left|\frac{\tilde b_2}{\tilde\alpha_4^u}\right| = 309^{+14}_{-17} \ ,  
\end{equation}
which numerically lifts part of the suppression from $C_{M_1M_2}^{B_q}$.
Further experimental input on these modes could further sharpen the picture and give insights into the annihilation modes. 

\subsubsection[$B_s^0\to \pi^+\pi^-$ and $B_d^0\to K^+K^-$]{\boldmath $B_s^0\to \pi^+\pi^-$ and $B_d^0\to K^+K^-$}
Finally, we comment on the pure annihilation modes $B_s^0\to \pi^+\pi^-$ and $B_d^0\to K^+K^-$, which only depend on $\tilde b_1 + 2\,\tilde b_4^u$ and their penguin counterparts. Current data already limits these parameters as seen in \eqref{eq:b1b4pen} and \eqref{eq:b1b4tree}, but the large uncertainties prevent any strong conclusions. 

The CP asymmetries of these two decays have not yet been measured.
From our analysis, we do not find any strong constraint for $B_d^0\to K^+K^-$ modes.
However, our analysis allows us to postdict:
\begin{equation}
    \mathcal{A}^{\rm dir}_{\rm CP}(B_s^0\to\pi^+\pi^-) = \left(-0.1_{-4.7}^{+4.6}\right)\% \ .
\end{equation}
Measurements of these CP asymmetries are highly anticipated to shed further light on the annihilation parameters and reduce their uncertainties.

\newpage
\section{Conclusion}
\label{sec:conclusion}

The phenomenology of hadronic two-body $B\to PP$ observables is very rich and offers a unique playground for understanding QCD at low scales.
A plethora of experimental data is available for different final states and for both branching ratios and CP asymmetries.
Yet, the theoretical description of these decays remains challenging. 

In this work, we performed a detailed analysis of $B\to PP$ decays, where $P=\pi, K$.
In light of the updated data, we first performed an analysis assuming \SUF-flavour symmetry.
Including mixing-induced CP asymmetries, correlations between modes and meson mixing, we find a poor description of the data, with a $p$-value below our threshold of $3\%$. 
The (expected) breakdown of \SUF symmetry can be seen already by considering only the $U$-spin partner decays $B_s^0\to K^+ K^-$ and $B^0\to \pi^+ \pi^-$.
Improved data in these channels, especially thanks to the recent observation of CP violation in the $B_s$ mode, already shows a violation of the \SUF-limit at the $2\sigma$ level.  

Going beyond the \SUF-limit, we incorporate factorizable \SUF-breaking effects stemming from form factors and decay constants.
To this end, we introduce a parametrisation similar to the standard QCD factorization parametrisation. 
We find that factorizable \SUF-breaking corrections give a perfect description of the data. 

More -- and more precise -- measurements of these modes will help sharpen the picture further and provide insight into the non-perturbative QCD effects. 
We have identified a number of key modes which benefit from our analysis.
These are specifically the $B_{d,s}^0\to K^0\bar{K}^0$ and $B^+\to K^0 K^+, B^+\to K^0 \pi^+$ modes, for which already updated branching ratio measurements would provide additional information.
In addition, measurements of the CP asymmetries in the annihilation modes, like $B\to K^+K^-$, would constrain the phases of the suppressed annihilation coefficients, which are notoriously problematic to calculate.
Updates of the -- already precisely measured -- $B_s^0\to K^+K^-$ mode would give insights into non-factorizable \SUF-breaking effects that are not included in our current analysis.
We highlight that we do not find any puzzling patterns in the $B\to \pi K$ decays as we perfectly accommodate the experimental data.
This may be understood from our finding that the data dictates that electroweak penguin parameters are of similar size as the QCD penguin parameters.
Despite our perfect description of the data, we can only constrain combinations of parameters with a satisfactory precision, such as $\alpha_1 + \alpha_4^u$ and $\alpha_2 - \alpha_4^u$.
Individual coefficients have broad distributions, making their comparison with calculations, \eg with QCD factorization, challenging.
A more detailed discussion of this is left for future work. 
Our nominal analysis does not include modes to $\eta^{(\prime)}$ final states.
Including factorizable \SUF-breaking effects in these modes requires a dedicated study, which we leave for future work~(see \cite{Beneke:2002jn} for a discussion within QCDF).
It would be interesting to extend this analysis to include also $B\to PV$ or even $B\to VV$, where $V=\rho, K^*$.
These modes introduce different dependencies on the parameters, which may break some of the poorly constrained direction we encountered.
However, we note that this requires a careful treatment of the finite width effects or going beyond a quasi-two-body approach (see e.g.~\cite{Krankl:2015fha, Klein:2017xti, Mannel:2020abt}). 
It will be interesting to see if factorizable \SUF-breaking continues to give a good description of the $B\to PP$ decays with improved data.
This suggests that \SUF-breaking is at the level of $20-30\%$, in line with general expectations.
We eagerly await new experimental results to further probe into QCD in these unique decays. 

\section*{Acknowledgments}
We are grateful to Danny van Dyk for valuable discussions.
The work of K.K.V. and M.B.M. is supported by the Dutch Research Council (NWO) as part of the project Solving Beautiful Puzzles (VI.Vidi.223.083) of the research programme Vidi.

\appendix

\newpage
\section{$\mathbf{A_\text{CP}^{\Delta\Gamma}}$ postdictions}\label{app:ADeltaGamma}
We provide our results for $A_{\rm CP}^{\Delta \Gamma}$ defined in \eqref{eq:AdelGam}. We note that this quantity is related to the direct and mixing-induced CP asymmetry given in Table~\ref{tab:table_topofit_Adir} and Table~\ref{tab:table_topofit_Amix} through the relation \cref{eq:SR_CP}. However, since the results for the direct and mixing-induced CP asymmetries are correlated and non-Gaussian, $A_{\rm CP}^{\Delta \Gamma}$ cannot be straightforwardly obtained using \eqref{eq:SR_CP} and the quoted median and uncertainty intervals. For completeness, we therefore provide our results here. 

\begin{table}[htp]
    \centering
    \scalebox{0.95}{
        \begin{tabular}{lccc}
          \toprule 
          \multirow{2}{*}{Channel} & \multicolumn{3}{c}{\textbf{$A_\text{CP}^{\Delta\Gamma}$ in units of $10^{-2}$}} \\[2pt]
          & \textbf{Experimental value} & \textbf{$\mathbf{\SUF}$} & \textbf{Fact.-$\cancel{\mathbf{\SUF}}$} \\
          \midrule
          ${B}^0\to \pi^+ \pi^-$& Not available &
          $60.4_{-2.1}^{+2.2}$ & $66.1 \pm 2.0$ \\[0.4em]
          ${B}^0\to \pi^0 \pi^0$& Not available &
          $10.3_{-9.5}^{+9.1}$ & $6^{+27}_{-30}$ \\[0.4em]
          ${B}^0\to K^+ K^-$& Not available &
          $22_{-16}^{+12}$ & $-29^{+30}_{-35}$ \\[0.4em]
          $B^0\to K^0 \bar{K}^0$& Not available &
          $-49 \pm 31$ & $22^{+59}_{-86}$ \\[0.4em]
          ${B}_s^0\to \pi^0 \bar K_S^0$& Not available &
          $-70.3_{-4.9}^{+5.8}$ & $-74^{+19}_{-10}$ \\[0.4em]
          \midrule
          ${B}^0\to \pi^0 {K}_S^0$ & Not available &
          $59.7 \pm 1.2$& $71.0^{+8.3}_{-11.1}$ \\[0.4em]
          ${B}_s^0\to \pi^+ \pi^-$& Not available &
          $-99.3_{-0.19}^{+0.22}$ & $-99.8_{-0.12}^{+0.16}$ \\[0.4em]
          ${B}_s^0\to \pi^0\pi^0$& Not available &
          $-99.3_{-0.19}^{+0.22}$ & $-99.8_{-0.12}^{+0.16}$ \\[0.4em]
          ${B}_s^0\to K^+ K^-$ & $-89.7 \pm 8.7$ \cite{LHCb:2020byh}&
          $-98.4 \pm 0.10$ & $-98.0^{+0.18}_{-0.15}$ \\[0.4em]
          ${B}_s^0\to K^0 \bar{K}^0$& Not available &
          $-99.86 \pm 0.12$ & $-85^{+31}_{-12}$ \\[0.4em]
          \bottomrule
        \end{tabular}}
     \caption{
        Experimental values and fit postdiction for the $A_\text{CP}^{\Delta \Gamma}$ observables.
        For our postdictions, we provide the medians and the central 68\% integrated probability intervals.
     }
     \label{tab:table_topofit_ADG}
\end{table}

\section{Postdictions for $\eta$ modes}\label{app:eta}

In \cref{tab:tableBr_combined,tab:tableAs_combined} we present the postdictions for decays involving $\eta$ mesons in the final state for our \SUF-limit analysis. We use that in the \SUF-limit, we have $|\eta\rangle = |\eta_8\rangle$. The amplitudes in terms of the topological parameters are given in \cite{He:2018php}. For comparison, we also quote the experimental data. 

\begin{figure}[ht]
    \centering
    \includegraphics[width=0.45\textwidth]{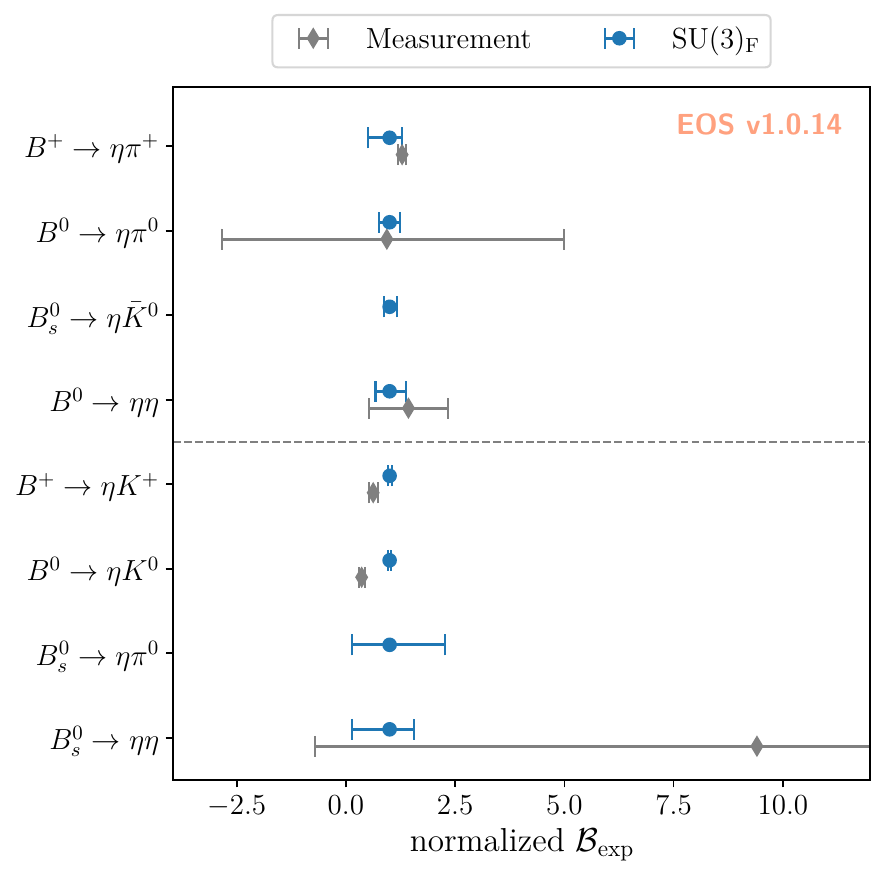} \hfill
    \includegraphics[width=0.45\textwidth]{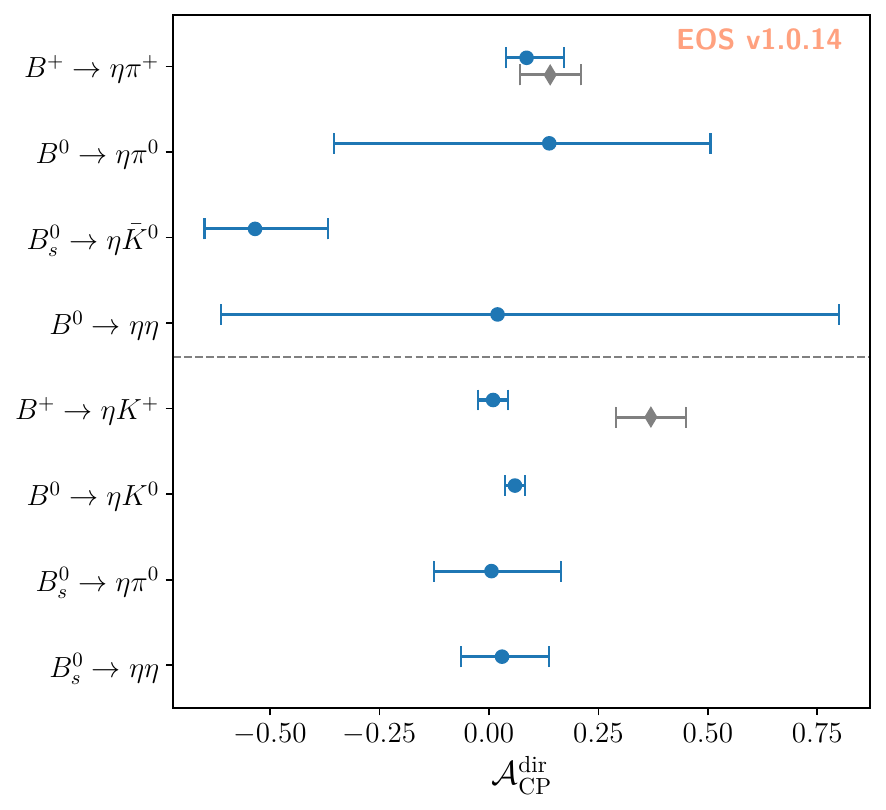}
    \caption{
        Predicted branching ratios and direct CP asymmetries of the decays involving $\eta$ mesons in the \SUF analysis.
        The observables are defined in \cref{sec:observables}.
        The few known measurements are in grey; they are not used in our fits.
    }
    \label{fig:Topofit_witheta}
\end{figure}

For completeness, we also show our postdictions and the available data in Fig.~\ref{fig:Topofit_witheta}.
We note that only 2 CP asymmetries are measured, both with sizeable uncertainties. 

We do not postdict these modes including factorizable \SUF-breaking as that would require a dedicated analysis.

\begin{table}[htp]
    \centering
    \scalebox{0.95}{
        \begin{tabular}{lcc}
            \toprule 
            \multirow{2}{*}{Channel} & \multicolumn{2}{c}{\textbf{Branching Ratios in units of $10^{-6}$}} \\[2pt]
            & \textbf{Experimental value} & \textbf{$\mathbf{\SUF}$} \\
            \midrule
            $B^+\to \eta \pi^+$& $4.02\pm 0.27$ & $3.1_{-1.6}^{+0.9}$ \\[0.4em]
            ${B}^0\to \eta \pi^0$& $0.41^{+0.17+0.05}_{-0.15-0.07}$\cite{Belle:2015eny} & $0.44 \pm 0.11$ \\[0.4em]
            ${B}_s^0\to \eta \bar K^0$&  Not available & $0.376_{-0.050}^{+0.063}$ \\[0.4em]
            ${B}^0\to \eta \eta$&$0.5 \pm 0.3 \pm 0.1$ \cite{BaBar:2009cun} & $0.35_{-0.11}^{+0.13}$ \\[0.4em]
            \midrule
            $B^+\to \eta K^+$&$2.4\pm 0.4$ & $3.84_{-0.17}^{+0.18}$\\[0.4em]   
            ${B}^0\to \eta K^0$&$1.23 {^{+0.27}_{-0.23}}$ & $3.4\pm 0.10$ \\[0.4em]
            ${B}_s^0\to \eta \pi^0$& $<10^{3}$ \cite{L3:1995lzt} & $11.4_{-9.9}^{+14.4}$ \\[0.4em]
            {${B}_s^0\to \eta \eta$}& {$100 \pm 105 \pm 23$}\cite{Belle:2021tsq} & $10.6_{-9.0}^{+5.9}$\\[0.4em]
            \bottomrule
        \end{tabular}
    }
    \caption{Experimental values and fit postdictions for $B\to \eta P$ branching ratios.
    Values without reference are from the PDG \cite{PDG2024}.}
    \label{tab:tableBr_combined}
\end{table}

\begin{table}[htp]
    \centering
    \scalebox{0.95}{
        \begin{tabular}{lcc}
            \toprule 
            \multirow{2}{*}{Channel} & \multicolumn{2}{c}{\textbf{Direct CP asymmetries in units of $10^{-2}$}} \\[2pt]
            & \textbf{Experimental value} & \textbf{$\mathbf{\SUF}$} \\
            \midrule
            $B^+\to \eta \pi^+$& $14\pm 7$ & $8.6_{-4.7}^{+8.6}$ \\[0.4em]
            ${B}^0\to \eta \pi^0$&Not available & $14_{-49}^{+36}$ \\[0.4em]
            ${B}_s^0\to \eta \bar K^0$&Not available & $-53_{-12}^{+17}$\\[0.4em]
            ${B}^0\to \eta \eta$&Not available & $2_{-63}^{+78}$\\[0.4em]
            \midrule
            $B^+\to \eta K^+$ & $37\pm 8$ & $1.0_{-3.4}^{+3.5}$ \\[0.4em]
            ${B}^0\to \eta K^0$& Not available & $6.0 \pm 2.3$\\[0.4em]
            ${B}_s^0\to \eta \pi^0$& Not available & $1_{-13}^{+16}$\\[0.4em]
            ${B}_s^0\to \eta \eta$& Not available & $3.0_{-9.5}^{+11.0}$ \\[0.4em]
            \bottomrule
         \end{tabular}
    }
    \caption{Experimental values and fit postdictions for $B\to \eta P$ direct CP asymmetries.
    Experimental values are from the PDG \cite{PDG2024}.}
    \label{tab:tableAs_combined}
\end{table}

\bibliography{refs}

\end{document}